\documentclass[journal]{IEEEtran}
\usepackage{xcolor}
\usepackage{hyperref}
\usepackage{tikz}
\usetikzlibrary{shapes.geometric, arrows}
\usetikzlibrary{patterns}
\tikzstyle{sq} = [rectangle, minimum width = 1.5cm, minimum height = 1.5cm, text centered, text width = 1.0cm, draw=black, fill=white!30]
\tikzstyle{arrow} = [thick,->,>=stealth]
\usepackage{pgfplots}
\usepgfplotslibrary{colorbrewer}
\usetikzlibrary{pgfplots.statistics, pgfplots.colorbrewer}
\usepackage{pgfplotstable}
\usepackage{comment}

\usepackage{cite}
\usepackage{amsmath}
\ifCLASSOPTIONcompsoc
  \usepackage[caption=false,font=normalsize,labelfont=sf,textfont=sf]{subfig}
\else
  \usepackage[caption=false,font=footnotesize]{subfig}
\fi
\usepackage{url}
\begin{document}
%
% paper title
% Titles are generally capitalized except for words such as a, an, and, as,
% at, but, by, for, in, nor, of, on, or, the, to and up, which are usually
% not capitalized unless they are the first or last word of the title.
% Linebreaks \\ can be used within to get better formatting as desired.
% Do not put math or special symbols in the title.
\title{FOVQA: Blind Foveated Video Quality Assessment}
%
%
% author names and IEEE memberships
% note positions of commas and nonbreaking spaces ( ~ ) LaTeX will not break
% a structure at a ~ so this keeps an author's name from being broken across
% two lines.
% use \thanks{} to gain access to the first footnote area
% a separate \thanks must be used for each paragraph as LaTeX2e's \thanks
% was not built to handle multiple paragraphs
%

\author{Yize Jin,
        Anjul Patney,
        Richard Webb,
        Alan C. Bovik,~\IEEEmembership{Fellow,~IEEE}% <-this % stops a space
\thanks{Y. Jin and A. C. Bovik are with the Department of Electrical and Computer Engineering, The University of Texas at Austin, Austin, TX, 78712 USA e-mail: yizejin@utexas.edu; bovik@ece.utexas.edu}% <-this % stops a space
\thanks{A. Patney was with Facebook Reality Labs. He is now with NVIDIA. e-mail: anjul.patney@gmail.com}
\thanks{R. Webb is with Facebook Reality Labs. e-mail: rwebb@fb.com}% <-this % stops a space
%\thanks{Manuscript received April 19, 2005; revised August 26, 2015.}
}
% note the % following the last \IEEEmembership and also \thanks - 
% these prevent an unwanted space from occurring between the last author name
% and the end of the author line. i.e., if you had this:
% 
% \author{....lastname \thanks{...} \thanks{...} }
%                     ^------------^------------^----Do not want these spaces!
%
% a space would be appended to the last name and could cause every name on that
% line to be shifted left slightly. This is one of those "LaTeX things". For
% instance, "\textbf{A} \textbf{B}" will typeset as "A B" not "AB". To get
% "AB" then you have to do: "\textbf{A}\textbf{B}"
% \thanks is no different in this regard, so shield the last } of each \thanks
% that ends a line with a % and do not let a space in before the next \thanks.
% Spaces after \IEEEmembership other than the last one are OK (and needed) as
% you are supposed to have spaces between the names. For what it is worth,
% this is a minor point as most people would not even notice if the said evil
% space somehow managed to creep in.

% The paper headers
\markboth{Journal of \LaTeX\ Class Files,~Vol.~14, No.~8, August~2015}%
{Shell \MakeLowercase{\textit{et al.}}: FOVQA: Blind Foveated Video Quality Assessment}
% The only time the second header will appear is for the odd numbered pages
% after the title page when using the twoside option.
% 
% *** Note that you probably will NOT want to include the author's ***
% *** name in the headers of peer review papers.                   ***
% You can use \ifCLASSOPTIONpeerreview for conditional compilation here if
% you desire.

% If you want to put a publisher's ID mark on the page you can do it like
% this:
%\IEEEpubid{0000--0000/00\$00.00~\copyright~2015 IEEE}
% Remember, if you use this you must call \IEEEpubidadjcol in the second
% column for its text to clear the IEEEpubid mark.

% use for special paper notices
%\IEEEspecialpapernotice{(Invited Paper)}

% make the title area
\maketitle

% As a general rule, do not put math, special symbols or citations
% in the abstract or keywords.
\begin{abstract}
Previous blind or No Reference (NR) video quality assessment (VQA) models largely rely on features drawn from natural scene statistics (NSS), but under the assumption that the image statistics are stationary in the spatial domain. Several of these models are quite successful on standard pictures. However, in Virtual Reality (VR) applications, foveated video compression is regaining attention, and the concept of space-variant quality assessment is of interest, given the availability of increasingly high spatial and temporal resolution contents and practical ways of measuring gaze direction. Distortions from foveated video compression increase with increased eccentricity, implying that the natural scene statistics are space-variant. Towards advancing the development of foveated compression / streaming algorithms, we have devised a no-reference (NR) foveated video quality assessment model, called FOVQA, which is based on new models of space-variant natural scene statistics (NSS) and natural video statistics (NVS). Specifically, we deploy a space-variant generalized Gaussian distribution (SV-GGD) model and a space-variant asynchronous generalized Gaussian distribution (SV-AGGD) model of mean subtracted contrast normalized (MSCN) coefficients and products of neighboring MSCN coefficients, respectively. We devise a foveated video quality predictor that extracts radial basis features, and other features that capture perceptually annoying rapid quality fall-offs. We find that FOVQA achieves state-of-the-art (SOTA) performance on the new 2D LIVE-FBT-FCVR database, as compared with other leading FIQA / VQA models. we have made our implementation of FOVQA available at: \url{http://live.ece.utexas.edu/research/Quality/FOVQA.zip}.
\end{abstract}

% Note that keywords are not normally used for peerreview papers.

% For peer review papers, you can put extra information on the cover
% page as needed:
% \ifCLASSOPTIONpeerreview
% \begin{center} \bfseries EDICS Category: 3-BBND \end{center}
% \fi
%
% For peerreview papers, this IEEEtran command inserts a page break and
% creates the second title. It will be ignored for other modes.
\IEEEpeerreviewmaketitle

\section{Introduction}
% The very first letter is a 2 line initial drop letter followed
% by the rest of the first word in caps.
% 
% form to use if the first word consists of a single letter:
% \IEEEPARstart{A}{demo} file is ....
% 
% form to use if you need the single drop letter followed by
% normal text (unknown if ever used by the IEEE):
% \IEEEPARstart{A}{}demo file is ....
% 
% Some journals put the first two words in caps:
% \IEEEPARstart{T}{his demo} file is ....
% 
% Here we have the typical use of a "T" for an initial drop letter
% and "HIS" in caps to complete the first word.
\IEEEPARstart{R}{ecent} advancements in Virtual Reality (VR) have drawn increasing attention to the development of immersive video contents, including high-resolution (4K+) $360^{\circ}$ videos. Until recently, head-mouted displays (HMDs) for VR have supported resolutions of about 1Kx1K to 2Kx2K. However, more recent HMDs deliver wide fields of view (FOV) approaching $200^{\circ}$, frame rates exceeding 75Hz, and the spatial resolutions that are approaching 8K. These spatial resolutions equate to angular resolutions of $10\sim 20$ pixels per degree (ppd), while the maximum resolution of the human eye approaches 120 ppd. Future immersive and $360^{\circ}$ video displays systems can benefit by increased resolutions which will drive even greater demand on the already significant bandwidth consumption.

One way to reduce bandwidth consumption is by using foveated protocols for video compression, which is a topic of increasing research interest, because of the availability of inexpensive and accurate consumer eyetrackers. Foveated compression techniques exploit the spatially decreasing acuity of the human vision system (HVS) away from the foveal center to achieve significant bandwidth savings, as for example by assigning larger quantization parameters (QP) to contents lying in the visual periphery. While several foveated compression algorithms \cite{Ryoo2016,Romero2018,Kim2018,Illahi2020} have been designed on top of modern video codec standards like H.264 / AVC and H.265 / HEVC, it is important to understand how the HVS perceives the outcomes of these compression protocols. For example, the authors of \cite{Illahi2020,Ryoo2016} conducted user studies to measure the quality of foveated compression / streaming algorithms they proposed. Towards creating a more generally applicable tools capable of predicting the perceptual quality of foveated / compressed contents, we recently designed 2D and 3D VR foveated video quality databases (LIVE-FBT-FCVR) \cite{yizetip2020} containing a wide spectrum of foveated and compressed distortions, on which we conducted extensive subjective studies of perceived quality. This new resource is intended to help escalate the development of accurate and efficient objective foveated video quality assessment (FVQA) models, which in turn can be used to help advance the development of improved foveated video compression techniques.

The field of objective FIQA / FVQA is sparse, especially in the area of no reference (NR) models. Although existing non-foveated (traditional) NR algorithms can be directly applied to foveated videos, they are unable to adequately capture the the perceptual effects of space-varying distortions. For example, even perceptually acceptable foveated videos may contain very low-quality contents in the visual periphery, as long as high-quality (low QP) contents fill the inner FOV, i.e., the foveal and parafoveal projections. Traditional algorithms way accurately respond to foveal and parafoveal distortions, but will inaccurately respond to (often intentional) degradations of peripheral contents, due to their underlying assumption of distortions that are uniformly distributed in the spatial domain.

Here we seek to advance progress on automatically assessing foveation distortions in the form of a new FVQA model, that we call FOVQA. FOVQA is driven by space-variant natural scene statistics (NSS) and natural video statistics (NVS) models \cite{dashi2020, lee2020}, wherein the assumptions of spatially stationarity are removed. Specifically, we deploy space-variant generalized Gaussian distribution (SV-GGD) and space-variant asynchronous generalized Gaussian distribution (SV-AGGD) models of the distributions of Mean Subtracted Contrast Normalized (MSCN) coefficients of VR videos having known or measured fixation coordinates. 

We also propose a number of other foveation-specific features which are able to capture important and unique factors that affect the perceptual quality of foveated videos. We advance progress towards this goal by designing a set of spatial weighting patterns to extract and pool features derived under SV-GGD and SV-AGGD over a discrete range of granularities. We also model an important perceptual phenomenon whereby foveated video quality is not only affected by the distortion levels, but also by the rapidity of quality fall-off from fovea to far periphery. We have found that this new source of significant quality degradation can be captured by analyzing the gradients of local video statistics.

We thoroughly tested the efficacy of FOVQA by a detailed ablation study, and comparing its performance against existing foveated and non-foveated VQA models. The rest of the paper is organized as follows: Section \ref{sect:s2} studies previous work on video quality assessment, including both traditional VQA models and foveated VQA models. Section \ref{sect:s3} describes the proposed FOVQA algorithm. Experiments and results are presented and discussed in Section \ref{sect:s4}, and Section \ref{sect:s5} concludes the paper and discusses possible future improvements.

\section{Related Work}\label{sect:s2}
Objective video quality assessment models have significantly evolved over the past two decades. Diverse application scenarios have guided researchers to develop models that rely on varying amounts of information from a pristine reference video, ranging from full reference (FR), reduced reference (RR), to no reference (NR) models.

FR models have been extensively studied and used, and are well exemplified by SSIM \cite{ssim} and MS-SSIM \cite{msssim}, whereby perceptually relevant luminance, contrast, and structure comparison measurements are integrated. The use of natural scene statistics models for picture quality prediction were first used in \cite{ifc,vif}, which model bandpass images as obeying a Gaussian scale mixture (GSM) \cite{gsm} model. The visual information fidelity (VIF) \cite{vif} deploys a neural noise model of uncertainty in the perceptual process. Another popular algorithm called the feature similarity (FSIM) index \cite{fsim} measures image phase congruency (PC) and gradient magnitude (GM) in a SSIM-like setting.

The aforementioned models, while often used to conduct VQA, do not make any temporal measurements. Among those that do, an early model called the Video Quality Metric (VQM) \cite{vqm} uses local spatial-temporal (S-T) features to predict video quality. The MOVIE index \cite{movie} models motion sensitive neural responses in extra-cortical area MT \cite{areamt} to extract temporal artifacts. The Video Multimethod Assessment Fusion (VMAF) \cite{vmaf} combines features from VIF \cite{vif}, Detail Loss Metric (DLM) \cite{dlm}, and frame differences, using them to train a Support Vector Regressor (SVR) to predict video quality.

RR VQA models are applicable in video quality monitoring scenarios, where only a small amount of information is drawn from the reference videos. Models like \cite{rred, strred, speedqa} exploit natural scene statistics (NSS) and natural video statistics (NVS) to measure distortion-induced statistical deviations of distorted videos from pristine videos.

Many existing NR VQA models rely on NSS and / or NVS. Frame-based algorithms like BRISQUE \cite{brisque} and NIQE\cite{niqe} extract simple spatial NSS parameters from bandpass and locally divisively normalized luminance frames, mapping them to quality predictions via an SVR \cite{svr} or a statistical distance \cite{niqe}. The Integrated Local NIQE (IL-NIQE) \cite{ilniqe} extends NIQE by incorporating gradient and chromatic statistics into the NIQE framework. V-BLIINDS \cite{vbliinds} injects temporal information into the video quality prediction process by employing natural video statistics (NVS) \cite{nvs} models of statistics of frame differences, and a motion masking model. VIIDEO \cite{viideo} extended NIQE by incorporating NVS into prediction without training. The Two Level Video Quality Model (TLVQM) \cite{tlvqm} takes a different approach by using a set of highly handcrafted features, obtaining SOTA performance on several datasets \cite{cvd2014,konvid1k,livequalcomm}.

Limited progress has been made on FIQA / FVQA models. The Foveated Wavelet Quality Index (FWQI) \cite{fwqi} combines an eccentricity-dependent contrast sensitivity function (CSF) \cite{csf} with a visually detectable noise threshold model \cite{vdntm}, to quantify the influence of peripheral distortions on the overall perceptual quality. The Foveated PSNR (FPSNR) and foveated weighted SNR (FWSNR) models \cite{fpsnr} account for foveated distortions by integrating curvilinear coordinate systems into the traditional PSNR / SNR metrics. The Foveation-based Content Adaptive SSIM (FA-SSIM) \cite{fassim} model combines SSIM with a foveation-based sensitivity function \cite{fovcsf}, whereby the effects of object velocity in the visual periphery are considered. A recent extension of BRISQUE to include foveation called Space-Variant BRISQUE (SVBRISQUE) \cite{sbrisque} deploys NSS and NVS models over foveation-graded concentric regions.

Towards further advancing progress on the foveated video quality prediction problem, we have developed a new prediction model called FOVQA that includes the following features:
\begin{itemize}
    \item We devised space-variant GGD and AGGD (SV-GGD and SV-AGGD) models, and use them to capture space-variant distortions that are characteristic of foveated compression. 
    \item We deploy a unique model of foveated quality fall-off, which we use to capture perceptual sensitivity to rapid changes in quality with increased eccentricity relative to visual fixations.
    \item Our feature extraction methods are all linear operations, hence simple online averaging can be used to stabilize the features or statistics computed from them across frames and viewing directions. This is much more memory efficient then accumulating video frames or coefficients computed from them.
\end{itemize}

\section{A Space-Variant NSS Model}\label{sect:s3}
\subsection{Statistics of Normalized Bandpass Coefficients}
In \cite{ruderman1994}, Ruderman pointed out that divisively normalizing bandpass-filtered natural images with the deviations of neighboring bandpass samples tends to yield decorrelated, Gaussian distributed coefficients. A simple version of the bandpass and normalization process is:

\begin{equation}
    \hat{I} = \frac{I(i,j)-\mu (i,j)}{\sigma (i,j) + C},
    \label{eq:MSCN}
\end{equation}
where $(i,j)$ are spatial indices, and $C$ is a stabilizing or saturation constant. The local mean $\mu$ and standard deviation $\sigma$ are:
\begin{equation}
    \mu (i,j) = \sum_{k=-K}^{K} \sum_{l=-L}^{L} w_{k,l}I_{k,l}(i,j)
    \label{eq:mu}
\end{equation}
and
\begin{equation}
    \sigma (i,j) = \sqrt{\sum_{k=-K}^{K}\sum_{l=-L}^{L}w_{k,l}(I_{k,l}(i,j)-\mu(i,j))^2},
    \label{eq:sigma}
\end{equation}

where $w$ is a 2D Gaussian weighting window of size $(2K+1,2L+1)$ sampled out to three standard deviations. We will refer to (\ref{eq:MSCN}) as mean subtracted contrast normalized (MSCN) coefficents. Widely-used NR VQA models seek to quantify perceptual distortions as a mapping between measurable distortions from these statistical regularities to perceptual quality \cite{sheikh2006,moorthy2010,brisque}. The empirical distributions (histograms) of the MSCN coefficients of both natural ($\alpha =2$) and distorted images can be modeled as following a generalized Gaussian distribution (GGD):
\begin{equation}
    f(x;\alpha,\sigma^2) = \frac{\alpha}{2\beta\Gamma(1/\alpha)}\exp{\left(-\left(\frac{|x|}{\beta}\right)^\alpha\right)},
    \label{eq:ggd}
\end{equation}
where $\alpha$ and $\sigma^{2}$ are shape and scale parameters,
\begin{equation}
    \beta = \sigma\sqrt{\frac{\Gamma(1/\alpha)}{\Gamma(3/\alpha)}},
    \label{eq:beta}
\end{equation}
and $\Gamma(\cdot)$ is the gamma function:
\begin{equation}
    \Gamma(a) = \int_{0}^{\infty}t^{a-1}e^{-t}dt \;\; a>0.
    \label{eq:gamma}
\end{equation}
Likewise, the products of pairs of adjacent MSCN coefficients have been effectively modeled as following zero mode asymmetric GGD (AGGD) models. The estimated parameters of both (\ref{eq:ggd}) and the AGGD models have been successfully used as quality-aware features. The paired products of MSCN coefficients are defined as:
\begin{equation}
    P(i,j) = \hat{I}(i,j)\hat{I}(i+d_{1},j+d_{2}),
    \label{eq:paird_prod}
\end{equation}
where $(d_{1}, d_{2}) \in \{(0,1),(1,0),(1,1),(1,-1)\}$, while the AGGD model is:
\begin{equation}
    f(x;\nu,\sigma_{l}^{2},\sigma_{r}^{2}) = 
    \begin{cases}
        \frac{\nu}{(\beta_{l}+\beta_{r})\Gamma(\frac{1}{\nu})}\exp\left(-\left(\frac{-x}{\beta_{l}}\right)^{\nu}\right)\;\;x<0\\
        \frac{\nu}{(\beta_{l}+\beta_{r})\Gamma(\frac{1}{\nu})}\exp\left(-\left(\frac{-x}{\beta_{r}}\right)^{\nu}\right)\;\;x\geq 0,
    \end{cases}
    \label{eq:aggd}
\end{equation}
where $\beta_{l}$ and $\beta_{r}$ are the scale parameters of the left half and the right half of (\ref{eq:aggd}).

\begin{figure*}[t]
    \centering
    \subfloat[Foveation Pattern]{\label{fig:featsmtx_pattern}\includegraphics[width=0.19\textwidth]{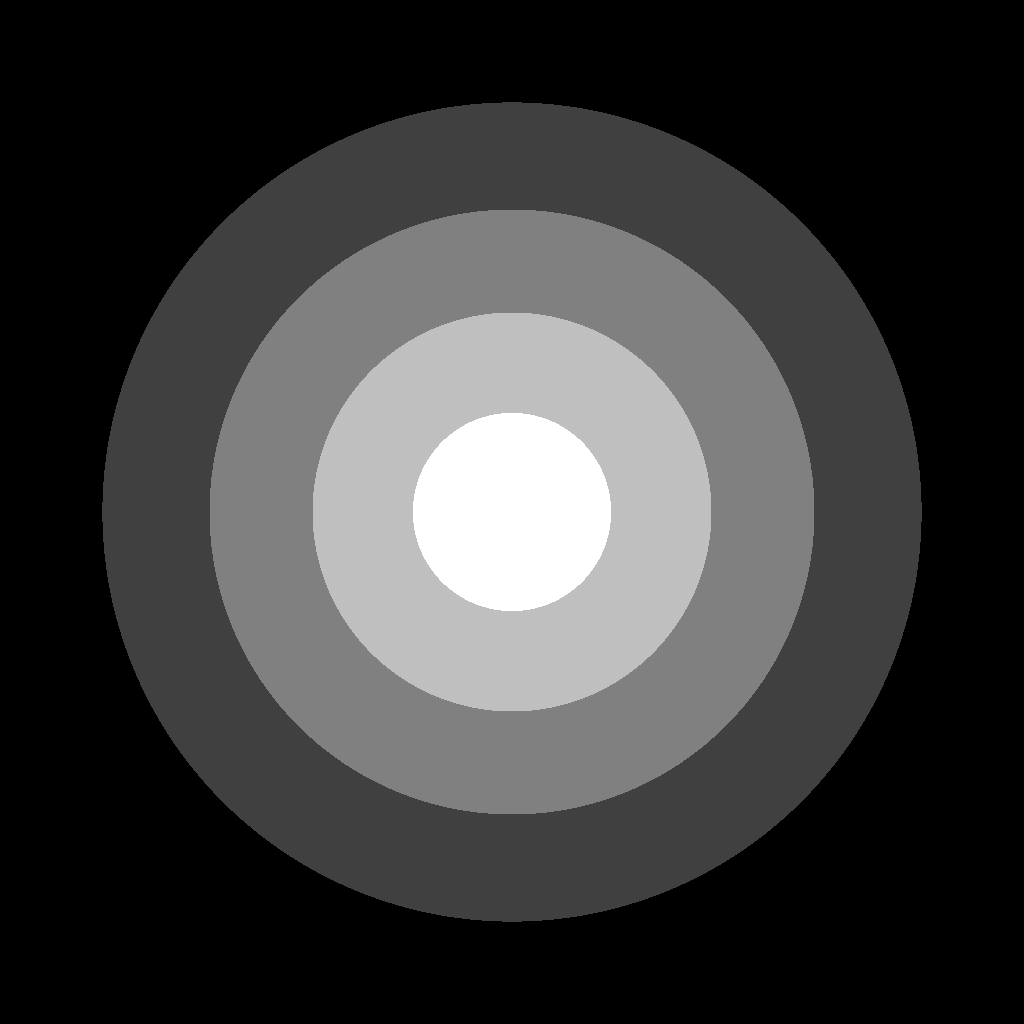}}\hfil
    \subfloat[Shape Matrix]{\label{fig:featsmtx_shape}\includegraphics[width=0.19\textwidth]{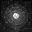}}\hfil
    \subfloat[Variance Matrix]{\label{fig:featsmtx_var}\includegraphics[width=0.19\textwidth]{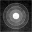}}\hfil
    \subfloat[Shape Matrix x0.5]{\label{fig:featsmtx_shape1}\includegraphics[width=0.19\textwidth]{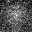}}\hfil
    \subfloat[Variance Matrix x0.5]{\label{fig:featsmtx_var1}\includegraphics[width=0.19\textwidth]{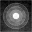}}
    \centering
    \caption{(a) Illustration of the 5-level foveation compression protocol. The compressions applied to the concentric regions (from center and moving outward) are no compression, $-crf=51$, $-crf=56$, $-crf=60$, and $-crf=63$, respectively, viz., increasing from cneter outward. (b) The spatial shape map of the local SV-GGD model of the foveated frame. (c) The variance map of the local SV-GGD model of the foveated frame. (d) The shape map of the SV-GGD model of the downscaled (by 2) foveated frame. (e) The variance map of the SV-GGD model of the downscaled foveated frame. It can be seen that (b), (c), (d) followed the foveation pattern in (a), while (d) is much more noisy.}
    \label{fig:featsmtx}
\end{figure*}

\subsection{Space-Variant GGD and AGGD Models}
Popular NSS / NVS based models like BRISQUE \cite{brisque} and V-BLIINDS \cite{vbliinds} generally assume that video distortions are uniformly distributed over space, hence MSCN coefficients are tacitly assumed to be spatially stationary. However, the distortions of foveated videos are intrinsically space-variant. To capture this property of foveated distortions, it is natural to model MSCN coefficients as instead following space-variant generalized Gaussian distributions (SV-GGD):
\begin{equation}
    \begin{aligned}
        \hat{I}(\mathbf{r}) &  \sim f(x;\alpha(\mathbf{r}),\sigma(\mathbf{r})^2)\\
        & = \frac{\alpha(\mathbf{r})}{2\beta(\mathbf{r})\Gamma(\frac{1}{\alpha(\mathbf{r})})}\exp{\left(-\left(\frac{|x|}{\beta(\mathbf{r})}\right)^{\alpha(\mathbf{r})}\right)},
    \end{aligned}
    \label{eq:svggd}
\end{equation}
where $\mathbf{r} = (i,j)$ are spatial indices, and $\alpha(\mathbf{r})$ and $\beta(\mathbf{r})$ vary spatially with $\mathbf{r}$. In a similar way, AGGD models can be made space-variant (SV-AGGD):
\begin{equation}
    \begin{aligned}
    & f(x;\nu(\mathbf{r}),\sigma_{l}^{2}(\mathbf{r}),\sigma_{r}^{2}(\mathbf{r})) = \\
    & \begin{cases}
        \frac{\nu(\mathbf{r})}{(\beta_{l}(\mathbf{r})+\beta_{r}(\mathbf{r}))\Gamma(\frac{1}{\nu(\mathbf{r})})}\exp\left(-\left(\frac{-x}{\beta_{l}(\mathbf{r})}\right)^{\nu(\mathbf{r})}\right)\;\;x<0\\
        \frac{\nu(\mathbf{r})}{(\beta_{l}(\mathbf{r})+\beta_{r}(\mathbf{r}))\Gamma(\frac{1}{\nu(\mathbf{r})})}\exp\left(-\left(\frac{-x}{\beta_{r}(\mathbf{r})}\right)^{\nu(\mathbf{r})}\right)\;\;x\geq 0,
    \end{cases}
    \end{aligned}
    \label{eq:svaggd}
\end{equation}
where $\nu(\mathbf{r})$, $\beta_{l}(\mathbf{r})$, and $\beta_{r}(\mathbf{r})$ also vary with $(\mathbf{r})$.

Estimation of the parameters ($\alpha, \beta, \nu,\beta_{l}, \beta_{r}$) requires special treatments. Since the maximum likelihood is, in this case, a functional of these space-variant parameters, maximum likelihood estimation (MLE) would require a difficult and expensive variational formulating. We do assume that the MSCN coefficients and their paired products are locally stationary, i.e. within a $P\times P$ window, the coefficients / products within share the same distribution, allowing the parameters to be estimated, e.g. by the popular moment-matching approach in \cite{sharifi1995}. In practice, we have found it sufficient and computationally efficient to partition each input frame into non-overlapping $P\times P$ patches, then estimate a set of parameters on each patch, without a loss of performance relative to a denser sampling. We also introduce a neural noise on the input images:
\begin{equation}
    \tilde{I}(i,j) = I(i,j) + \mathcal{W}_s,
    \label{eq:noise}
\end{equation}
where $\mathcal{W}_s \sim \mathcal{N}(0,\sigma_{w_s}^{2})$, $I(i,j)\in [0,255]$ and the modified MSCN coefficients are calculated as:
\begin{equation}
    \hat{I} = \frac{\tilde{I}(i,j)-\tilde{\mu} (i,j)}{\tilde{\sigma} (i,j) + C_s}.
    \label{eq:MSCN_noise}
\end{equation}

The neural noise model is motivated in two ways. First, similar to \cite{vif}, it is a way for accounting for uncertainty of visual perception, including noise affecting neurons along the visual pathway. Unlike \cite{vif}, the noise is not hypothetical, since we explicitly introduce small amounts of simulated noise on the video frames before processing via (\ref{eq:MSCN_noise}). This has the second important benefit of introducing a small amount of variation on constant or near-constant frame regions, sewing to regulate the behaviour of (\ref{eq:MSCN_noise}) where numeric zeros may occur, as on saturated over- or under-exposed portions of frames. These kinds of imperfections often occur on videos captured by current $360^{\circ}$ videos because of their (typically) limited dynamic ranges.

As a result, we obtain an $\lfloor{M/P}\rfloor \times \lfloor{N/P}\rfloor$ matrix of each parameter, where $(M,N)$ indicate the resolution of the input image, and each element of the matrix is a quality-aware feature corresponding to a local image patch.

\subsection{Radial Basis Feature Extraction} \label{sect:weight_patt}
While NSS features have proven to be highly predictive of perceptual video quality, current models apply them under the stationarity assumption. If applied to nonstationary, foveated videos, they are also nonstationary, which must be accounted for within any VQA model utilizing them.

As shown Fig. \ref{fig:featsmtx}, we begin by synthesizing a 5-level foveation distortion by first dividing the field of view (FOV) of a given video frame into 5 concentric regions. Our method of foveation involves compressing the content in each of the annular regions using different quantization parameters (QPs), as shown in Fig. \ref{fig:featsmtx_pattern}. An alternative would be to blur, then compress each region, but we have found that this added complexity does not improve results. To test this idea, we sampled 100 video frames from among the contents in \cite{yizetip2020} with this foveated compression protocal applied with various parameters, each of resolution 1024x1024 and a field of view (FOV) of $90^{\circ}$. On each of these foveation distorted video viewports, we computed the MSCN coefficients using equations \ref{eq:noise} and \ref{eq:MSCN_noise} with $\sigma_{w_{s}} = 10^{-2}$, and $C_{s} = 0.1$. We then estimated the parameter maps of the best fitting SV-GGD and SV-AGGD models by setting $P=32$ on each foveated frame and averaged the per-frame parameter maps. We plotted the averaged spatial shape and variance maps of the best-fitting SV-GGD model of the MSCN coefficents of the foveated viewports in Figs. \ref{fig:featsmtx_shape} and \ref{fig:featsmtx_var}, respectively, then downscaled the foveated frame by 2, and again plotted the averaged shape and variance maps in Figs. \ref{fig:featsmtx_shape1} and \ref{fig:featsmtx_var1}, respectively. It can be seen that the shape and variance maps of the frame followed the foveation pattern in Fig. \ref{fig:featsmtx_pattern}. However, on the downscaled viewports, while the variance map followed the foveation pattern, the shape map is quite noisy. Hence, we did not use the shape map from downscaled video viewports. We also discarded the shape map of the SV-AGGD model of the downscaled viewports, for the same reason. We observed similar noisy patterns of the local shape parameters of downscaled frames from the LIVE IQA and VQA databases \cite{sheikh2006, livevqa}, as well, showing that this phenomenon generally unlikely to be a database bias.

\begin{figure}[t]
    \centering
    \subfloat[]{\label{fig:dcg1}\includegraphics[width=0.45\columnwidth]{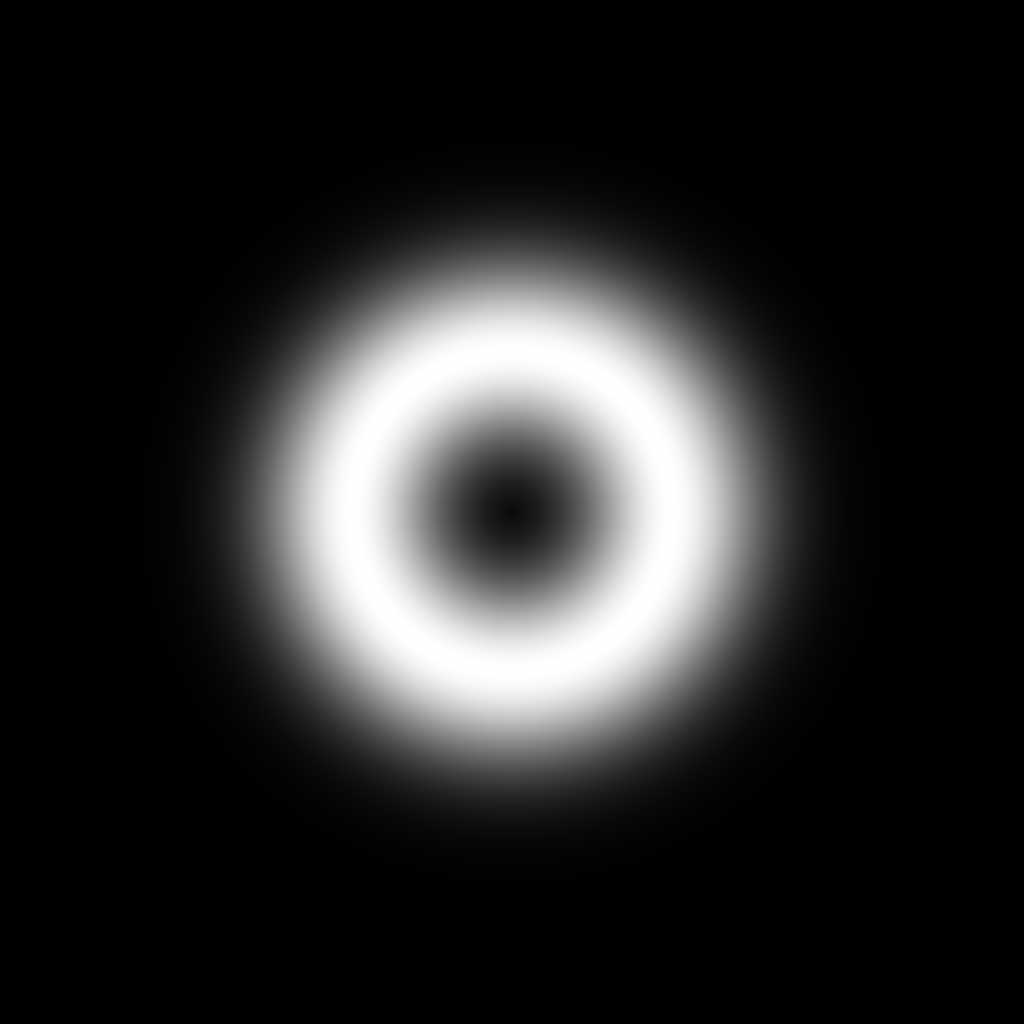}}\hfil
    \subfloat[]{\label{fig:dcg2}\includegraphics[width=0.45\columnwidth]{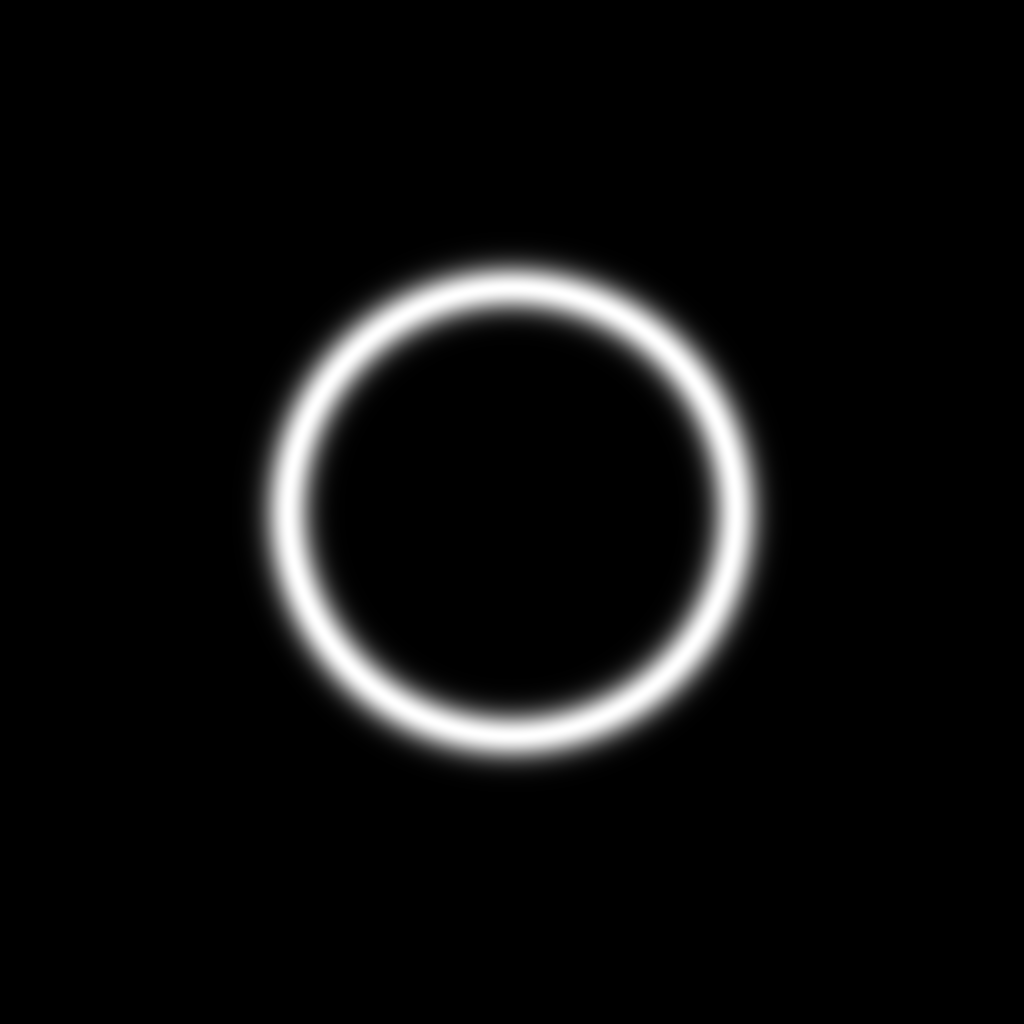}}\hfil
    \centering
    \caption{Examples of toroidal functions of differing radii $r_{k}$ and widths $\sigma_{G_{RBF}}$. Different choices of $r_{k}$ and $\sigma_{G_{RBF}}$ makes it possible to extract quality-aware NSS information at different granularities and eccentricities.}
    \label{fig:dcg}
\end{figure}

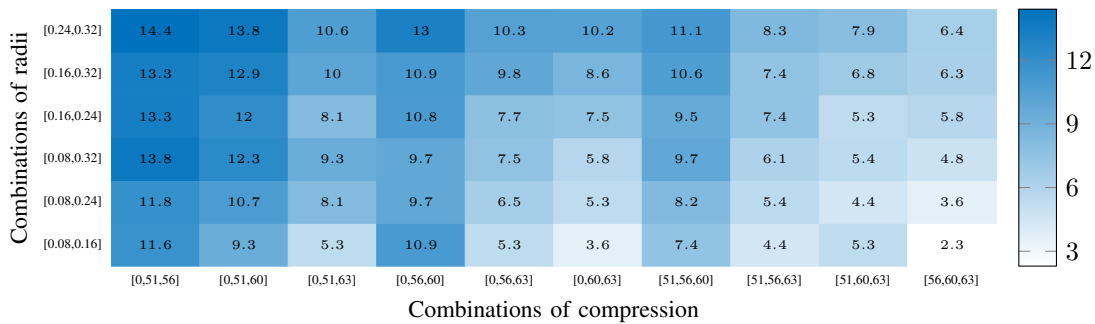
\begin{figure*}[t]
    \centering
    \pgfplotsset{width=0.92*0.8\textwidth,colormap={heatmap}{rgb255=(255,255,255) rgb255=(0, 114, 189)}}
    \begin{tikzpicture}
    \centering
    \begin{axis}[colormap name=heatmap,
                nodes near coords,
                enlargelimits=false,
                x label style={at={(0.5,-0.1)},anchor=south, font=\small},
                y label style={at={(0.02,0.5)},anchor=south, font=\small},
                xlabel = {Combinations of compression},
                ylabel = {Combinations of radii},
                colorbar,
                colorbar style={ytick={3,6,9,12}},
                height=5cm,
                axis line style={draw=none},
                xtick=\empty,
                ytick=\empty,
                every node near coord/.style={
                font=\tiny, % smaller text size, bold for the data point labels
                text=black,
                anchor=center % center the labels on the plot marks
                }]
        \addplot[matrix plot*,point meta=explicit] table[x=x, y=y, meta=val] {figures/map_2D_new.txt};
    \end{axis}
    \node[below] at (0.525*0.914/0.95/0.7*0.815,0) {\tiny{[0,51,56]}};
    \node[below] at (1.57*0.914/0.95/0.7*0.815,0) {\tiny{[0,51,60]}};
    \node[below] at (2.62*0.914/0.95/0.7*0.815,0) {\tiny{[0,51,63]}};
    \node[below] at (3.67*0.914/0.95/0.7*0.815,0) {\tiny{[0,56,60]}};
    \node[below] at (4.72*0.914/0.95/0.7*0.815,0) {\tiny{[0,56,63]}};
    \node[below] at (5.77*0.914/0.95/0.7*0.815,0) {\tiny{[0,60,63]}};
    \node[below] at (6.82*0.914/0.95/0.7*0.815,0) {\tiny{[51,56,60]}};
    \node[below] at (7.87*0.914/0.95/0.7*0.815,0) {\tiny{[51,56,63]}};
    \node[below] at (8.92*0.914/0.95/0.7*0.815,0) {\tiny{[51,60,63]}};
    \node[below] at (9.97*0.914/0.95/0.7*0.815,0) {\tiny{[56,60,63]}};
    \node[left] at (0,0.285) {\tiny{[0.08,0.16]}};
    \node[left] at (0,0.855) {\tiny{[0.08,0.24]}};
    \node[left] at (0,1.425) {\tiny{[0.08,0.32]}};
    \node[left] at (0,1.995) {\tiny{[0.16,0.24]}};
    \node[left] at (0,2.565) {\tiny{[0.16,0.32]}};
    \node[left] at (0,3.135) {\tiny{[0.24,0.32]}};
    \end{tikzpicture}
    \vspace{-3mm}
    \caption{Mean Ranked Opinion Scores (MROS) of each foveation distortion in the 2D LIVE-FBT-FCVR database, represented in combinations of compression and radii. Higher MROS indicates that the combination / distortion is considered to have higher perceptual quality. By comparing $[0,51,63]$ (the $3^{rd}$ column) against $[0,56,60]$ (the $4^{th}$ column), and $[0,56,63]$ (the $5^{th}$ column) against $[51,56,60]$ (the $7^{th}$ column), where the changes in foveation distortion occur smoothly rather than with similar compression but sharper changes of distortion, smoother quality fall-offs from the center of foveation generally lead to better perceptual quality.}
    \label{fig:falloff}
\end{figure*}

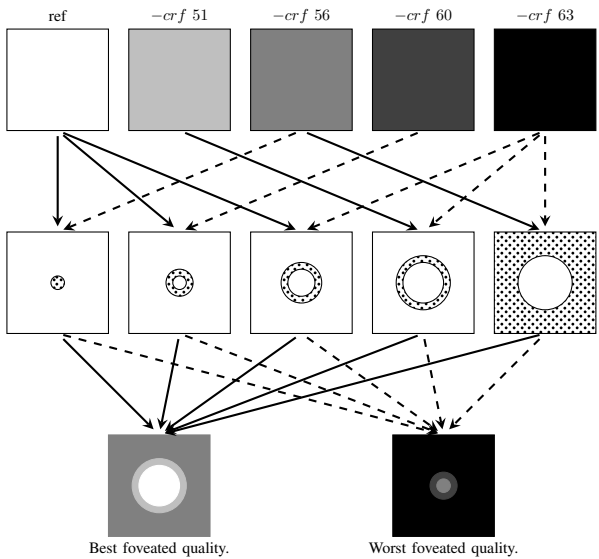
\begin{figure}[t]
    \centering
    \def\unit{0.9}
    \def\colorone{white}
    \def\colortwo{black}
    \begin{tikzpicture}[node distance=2cm,every node/.style={scale=0.6}]
    %draw ``compressed" images.
    \draw[fill=\colorone] (0,0) rectangle (1.5*\unit,1.5*\unit);
    \node at (0.75*\unit, 1.7*\unit) {ref};
    \node (level1) at (0.75*\unit, 0) {};
    \draw[fill=\colorone!75!\colortwo] (1.8*\unit,0*\unit) rectangle (3.3*\unit,1.5*\unit);
    \node at (2.55*\unit, 1.7*\unit) {$-crf\ 51$};
    \node (level2) at (2.55*\unit, 0) {};
    \draw[fill=\colorone!50!\colortwo] (3.6*\unit,0*\unit) rectangle (5.1*\unit,1.5*\unit);
    \node at (4.35*\unit, 1.7*\unit) {$-crf\ 56$};
    \node (level3) at (4.35*\unit, 0) {};
    \draw[fill=\colorone!25!\colortwo] (5.4*\unit,0*\unit) rectangle (6.9*\unit,1.5*\unit);
    \node at (6.15*\unit, 1.7*\unit) {$-crf\ 60$};
    \node (level4) at (6.15*\unit, 0) {};
    \draw[fill=\colortwo] (7.2*\unit,0*\unit) rectangle (8.7*\unit,1.5*\unit);
    \node at (7.95*\unit, 1.7*\unit) {$-crf\ 63$};
    \node (level5) at (7.95*\unit, 0) {};
    
    %draw ``radii".
    \draw (0, -1.5*\unit) rectangle (1.5*\unit, -3*\unit);
    \draw[pattern=crosshatch dots] (0.75*\unit, -2.25*\unit) circle (0.1*\unit);
    \node (r1_1) at (0.75*\unit, -1.5*\unit) {};
    \node (r1_2) at (0.75*\unit, -3*\unit) {};
    \draw (1.8*\unit, -1.5*\unit) rectangle (3.3*\unit, -3*\unit);
    \draw[pattern=crosshatch dots] (2.55*\unit, -2.25*\unit) circle (0.2*\unit);
    \draw[fill=white] (2.55*\unit, -2.25*\unit) circle (0.1*\unit);
    \node (r2_1) at (2.55*\unit, -1.5*\unit) {};
    \node (r2_2) at (2.55*\unit, -3*\unit) {};
    \draw (3.6*\unit, -1.5*\unit) rectangle (5.1*\unit, -3*\unit);
    \draw[pattern=crosshatch dots] (4.35*\unit, -2.25*\unit) circle (0.3*\unit);
    \draw[fill=white] (4.35*\unit, -2.25*\unit) circle (0.2*\unit);
    \node (r3_1) at (4.35*\unit, -1.5*\unit) {};
    \node (r3_2) at (4.35*\unit, -3*\unit) {};
    \draw (5.4*\unit, -1.5*\unit) rectangle (6.9*\unit, -3*\unit);
    \draw[pattern=crosshatch dots] (6.15*\unit, -2.25*\unit) circle (0.4*\unit);
    \draw[fill=white] (6.15*\unit, -2.25*\unit) circle (0.3*\unit);
    \node (r4_1) at (6.15*\unit, -1.5*\unit) {};
    \node (r4_2) at (6.15*\unit, -3*\unit) {};
    \draw[pattern=crosshatch dots] (7.2*\unit, -1.5*\unit) rectangle (8.7*\unit, -3*\unit);
    \draw[fill=white] (7.95*\unit, -2.25*\unit) circle (0.4*\unit);
    \node (r5_1) at (7.95*\unit, -1.5*\unit) {};
    \node (r5_2) at (7.95*\unit, -3*\unit) {};
    
    % draw ``foveated" images.
    \draw[fill=\colorone!50!\colortwo,\colorone!50!\colortwo] (1.5*\unit, -4.5*\unit) rectangle (3*\unit, -6*\unit);
    \draw[fill=\colorone!75!\colortwo, \colorone!75!\colortwo] (2.25*\unit, -5.25*\unit) circle (0.4*\unit);
    \draw[fill=\colorone, \colorone] (2.25*\unit, -5.25*\unit) circle (0.3*\unit);
    \node (fov_1) at (2.25*\unit, -4.5*\unit) {};
    \draw[fill=\colortwo] (5.7*\unit, -4.5*\unit) rectangle (7.2*\unit, -6*\unit);
    \draw[fill=\colorone!25!\colortwo, \colorone!25!\colortwo] (6.45*\unit, -5.25*\unit) circle (0.2*\unit);
    \draw[fill=\colorone!50!\colortwo, \colorone!50!\colortwo] (6.45*\unit, -5.25*\unit) circle (0.1*\unit);
    \node (fov_2) at (6.45*\unit, -4.5*\unit) {};
    \node[below] at (2.25*\unit, -6*\unit) {Best foveated quality.};
    \node[below] at (6.45*\unit, -6*\unit) {Worst foveated quality.};
    
    % draw arrows.
    \draw[arrow] (level1) -- (r1_1);
    \draw[arrow] (level1) -- (r2_1);
    \draw[arrow] (level1) -- (r3_1);
    \draw[arrow] (level2) -- (r4_1);
    \draw[arrow] (level3) -- (r5_1);
    \draw[arrow] (r1_2) -- (fov_1);
    \draw[arrow] (r2_2) -- (fov_1);
    \draw[arrow] (r3_2) -- (fov_1);
    \draw[arrow] (r4_2) -- (fov_1);
    \draw[arrow] (r5_2) -- (fov_1);
    \draw[arrow, dashed] (level3) -- (r1_1);
    \draw[arrow, dashed] (level4) -- (r2_1);
    \draw[arrow, dashed] (level5) -- (r3_1);
    \draw[arrow, dashed] (level5) -- (r4_1);
    \draw[arrow, dashed] (level5) -- (r5_1);
    \draw[arrow, dashed] (r1_2) -- (fov_2);
    \draw[arrow, dashed] (r2_2) -- (fov_2);
    \draw[arrow, dashed] (r3_2) -- (fov_2);
    \draw[arrow, dashed] (r4_2) -- (fov_2);
    \draw[arrow, dashed] (r5_2) -- (fov_2);
    
    %\node (rect1) [rct, minimum width = 1.5*\unit, minimum height=0.5*\unit] {ref};
    \end{tikzpicture}
    \vspace{-3mm}
    \caption{Ilustration of methods of creating foveated / compressed videos. The inner radii define concentric regions as shown in the second row. The distorted videos are defined by selecting inner radii that separate the multiple adjacent foveation regions. The solid arrows indicate the best foveated quality that is allowed, while the dashed arrows indicate the worst foveated quality possible.}
    \label{fig:fov_dist}
\end{figure}

Following these observations and building upon our preliminary ideas \cite{sbrisque}, we employed a set of normalized weighting patterns to capture foveation-specific features from the parameter maps. First define a set of toroidal functions indexed $k \in {1,2,...,\mathcal{K}}$ by convolving isotropic (Dirac) impulse rings of radii $r_{k}$ with a Gaussian function:
\begin{equation}
\begin{split}
    w_{k}(\mathbf{r}) &= \left(\frac{1}{|\sum_{\mathbf{r}}w_{k}(\mathbf{r})|}\right)\cdot\left[\delta(||\mathbf{r}||-r_{k})*G(||\mathbf{r}||; \sigma_{G_{RBF}})\right],
    \label{eq:weightpatt}
\end{split}
\end{equation}
where $\delta(||\mathbf{r}||)$ and $*$ indicate Dirac function and the convolution operation, respectively, and where $\mathbf{r} = (i,j)$ indicates spatial indices, with $i \in \{1,2,...,\lfloor{M/P}\rfloor\}, j \in \{1,2,...,\lfloor{N/P}\rfloor\}$. The parameter $\sigma_{G_{RBF}}$ dictates the spread of $G(||\mathbf{r}||; \sigma_{G_{RBF}})$, and hence the widths of the toroidal functions, making it possible to extract quality-aware NSS at various granularities and eccentricities, as illustrated in Fig. \ref{fig:dcg}. The norm $||\mathbf{r}||$ is defined by:
\begin{equation}
    %||\mathbf{x}||^{2} = \frac{(i-\frac{\lfloor{M/P}\rfloor+1}{2})^{2}}{(\frac{\lfloor{M/P}\rfloor}{\lfloor{N/Q}\rfloor})^{2}} + (j-\frac{\lfloor{N/Q}\rfloor+1}{2})^{2}.
    ||\mathbf{r}||^{2} = (i-\frac{\lfloor{M/P}\rfloor+1}{2})^{2} + (j-\frac{\lfloor{N/P}\rfloor+1}{2})^{2},
\end{equation}
where we assume the gaze point is the center of the image. We extract / pool $\mathcal{K}$ features from each parameter map by weighted summation:
\begin{equation}
    f_{k,m} = \sum_{\mathbf{x}} w_{k}(\mathbf{r})\cdot p_{m}(\mathbf{r}),
    \label{eq:rb_feats}
\end{equation}
where $p_{m}$ indicates the $m^{th}$ parameter matrix. 

In our implementation, first estimate the shape ($\alpha(\mathbf{r})$) and variance maps ($\beta(\mathbf{r})$) from each input frame under the SV-GGD model (\ref{eq:svggd}), then estimate one shape map ($\nu(\mathbf{r})$), one mean map \cite{brisque, prafulxray}, and two variance maps ($\beta_{l}(\mathbf{r}), \beta_{r}(\mathbf{r})$) for each of the four products of adjacent pairs in (\ref{eq:paird_prod}) modeled as SV-AGGD (\ref{eq:svaggd}), yielding 18 feature maps, thus $18\mathcal{K}$ features are extracted using (\ref{eq:rb_feats}). Then, from the downscaled input, only estimate a variance map from the SV-GGD model and two variance maps for each of the four paired products modeled as SV-AGGD, obtaining 9 additional feature maps, hence $9\mathcal{K}$ extracted features. Overall, we obtain $27\mathcal{K}$ features from the feature extraction process. In our experiments and in the final FOVQA model, $\mathcal{K} = 10$, hence 270 features are learned from overall.
\subsection{Modelling Rapid Quality Fall-off}
An interesting phenomenon that we observed on the LIVE-FBT-FCVR databases is that rapid quality fall-offs with increasing eccentricity appear to cause greater degradations of perceptual quality, even if the intrinsic distortion is less. On the LIVE-FBT-FCVR databases, foveation distortions were created by i) first sampling the space of compressed videos using 4 QP values ($-crf = 51,56,60,63$ in VP9), yielding 5 levels of perceptually discriminable levels of uniform distortions (including the references), ii) then dividing the FOV into one central, three annular, and one peripheral region, (hence 4 radii $0.08,0.16,0.24,0.32$ in radians), iii) and finally choosing combinations of 3 of 5 compression levels, and 2 of 4 radii, as shown in Fig. \ref{fig:fov_dist}.

We illustrate the influence of rapid quality fall-offs in Fig. \ref{fig:falloff}, wherein Mean Ranked Opinion Scores (MROS) were calculated on each foveated / distorted video created by combinations of compression and radii. Higher MROS indicates that the perceptual quality of the combination was more preferred by subjects. It may be observed that generally, MROS took higher values near the upper-left corner, where less compression distortion was introduced and larger radii were used (larger areas of high quality contents), as might be expected. However, this does not explain the relatively high MROS given to the compression combinations $[0,56,60]$ (the $4^{th}$ column) and $[51,56,60]$ (the $7^{th}$ column), which represent smoother quality fall-offs than the neighboring compression combinations / columns. Moreover, by comparing different combinations of radii, it may be observed that the differences of MROS between smooth combinations and their neighbors become more evident when the radii are smaller, suggesting that any quality fall-off effects have a reciprocal relationship with eccentricity, viz., nearer the fovea.

\begin{figure*}
    \captionsetup{justification=centering}
    \centering
    \subfloat{\label{fig:1}\includegraphics[width=0.19\textwidth]{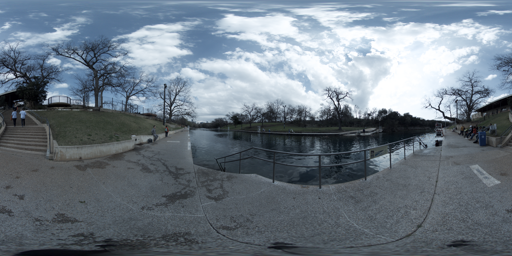}}\hfil
    \subfloat{\label{fig:2}\includegraphics[width=0.19\textwidth]{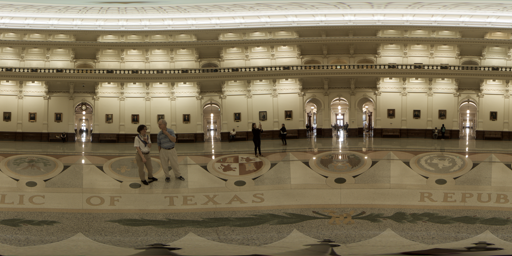}}\hfil
    \subfloat{\label{fig:3}\includegraphics[width=0.19\textwidth]{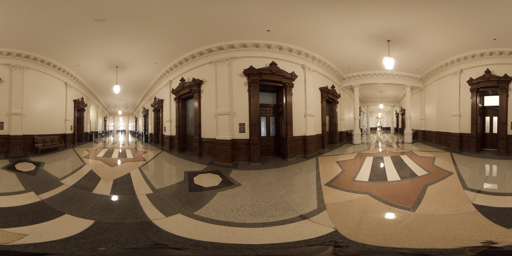}}\hfil
    \subfloat{\label{fig:4}\includegraphics[width=0.19\textwidth]{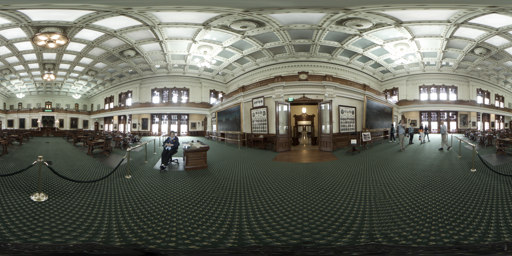}}\hfil
    \subfloat{\label{fig:5}\includegraphics[width=0.19\textwidth]{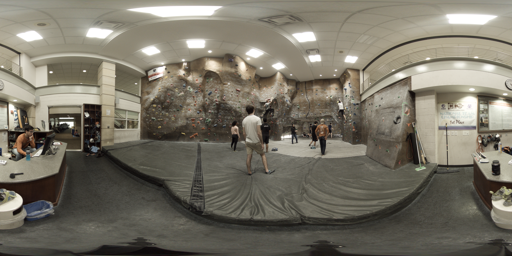}}\hfil\\
    \subfloat{\label{fig:6}\includegraphics[width=0.19\textwidth]{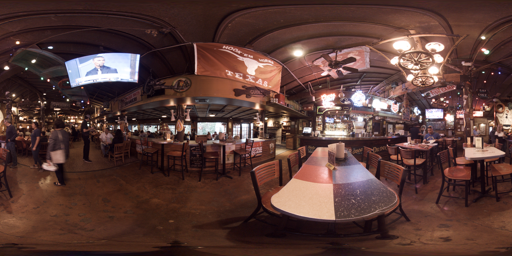}}\hfil
    \subfloat{\label{fig:7}\includegraphics[width=0.19\textwidth]{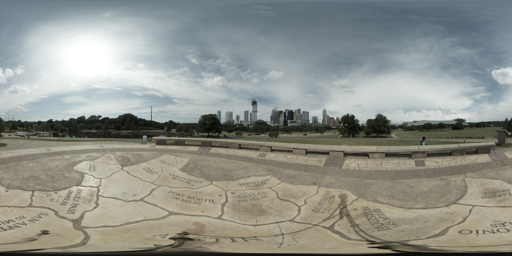}}\hfil
    \subfloat{\label{fig:8}\includegraphics[width=0.19\textwidth]{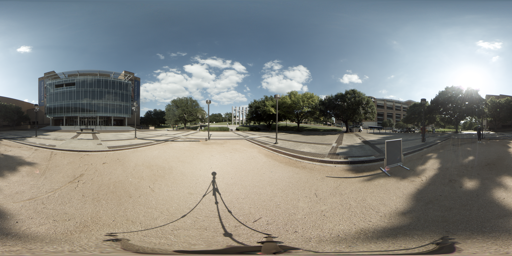}}\hfil
    \subfloat{\label{fig:9}\includegraphics[width=0.19\textwidth]{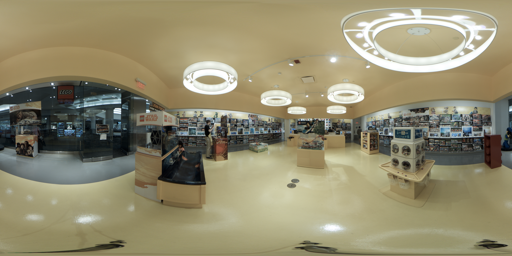}}\hfil
    \subfloat{\label{fig:10}\includegraphics[width=0.19\textwidth]{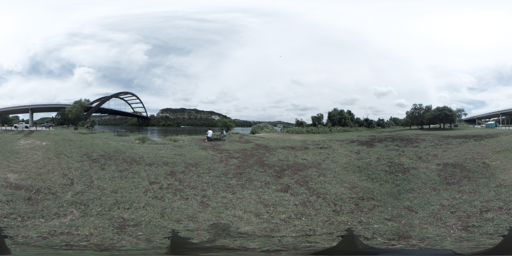}}\hfil\\
    \subfloat{\label{fig:11}\includegraphics[width=0.19\textwidth]{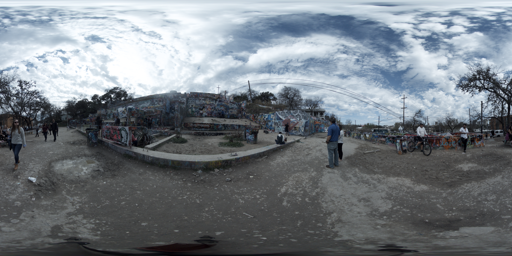}}\hfil
    \subfloat{\label{fig:12}\includegraphics[width=0.19\textwidth]{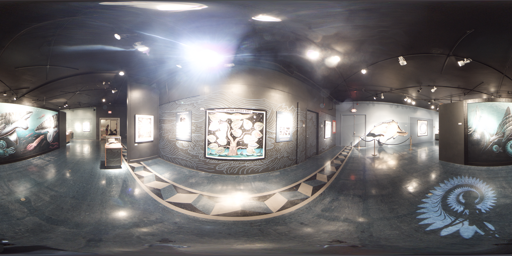}}\hfil
    \subfloat{\label{fig:13}\includegraphics[width=0.19\textwidth]{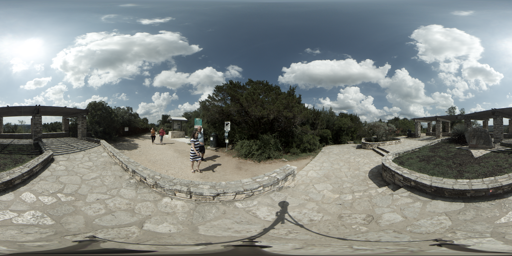}}\hfil
    \subfloat{\label{fig:14}\includegraphics[width=0.19\textwidth]{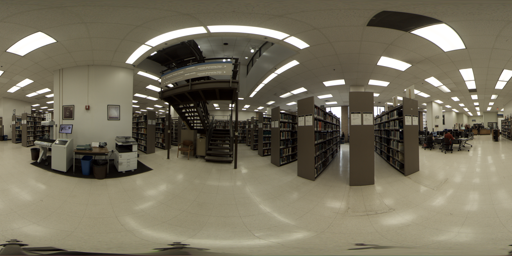}}\hfil
    \subfloat{\label{fig:15}\includegraphics[width=0.19\textwidth]{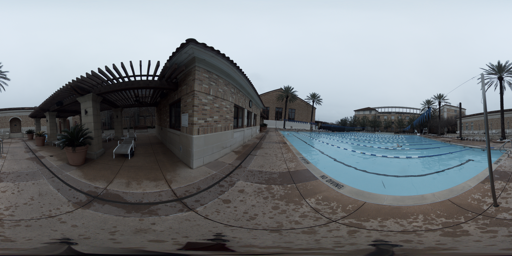}}\hfil\\
    \centering
    \caption{The QP prediction database. For all 15 pristine images, each was compressed into 6 distorted versions via VP9 using QP values $21,42,51,56,60$, and $63$, then 18 viewport images were generated for each of the 7 versions (including the reference). Then 3 NSS features were computed on each viewport image, and averaged across all the viewports. Finally an SVR was trained to map the features to QP values.}
    \label{fig:NSS2QP}
\end{figure*}

\begin{figure}
    \captionsetup{justification=centering}
    \centering
    \subfloat[]{\label{fig:qpref}\includegraphics[width=0.45\columnwidth]{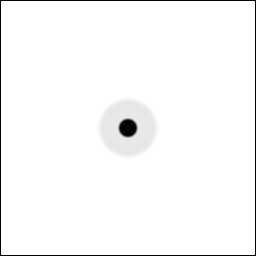}}\hfil
    \subfloat[]{\label{fig:qppred}\includegraphics[width=0.45\columnwidth]{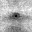}}\hfil
    \centering
    \caption{(a) A ground truth QP map. (b) The corresponding predicted QP map averaged across viewports and frames. It may be seen that the predicted QP map follows the ground truth map, but is noisier as could be expected.}
    \label{fig:QP_maps_comp}
\end{figure}

Based on these observations, we devised a model of the effects of quality fall-off. Sudden spatial increases in compression can be quite noticeable, especially on moving content. Fall-off can be captured by computing the gradient information on the ground-truth QP map of the foveated frame. However, in the blind video quality prediction setup, information regarding the QP values is usually not available. Hence, we first designed a database on which a QP predictor can be learned to map NSS features to QP values, then extracted the gradient information in the predicted QP maps.

We first collected 15 8K pristine immersive images which are different from the contents in the LIVE-FBT-FCVR databases, all of them in ERP format, as shown in Fig. \ref{fig:NSS2QP}. Then, we compressed these images using VP9 into 6 distorted versions, using $-crf=21,42,51,56,60,$ and $63$. To better simulate VR viewing condition, we sampled 18 3D viewing directions on a unit sphere, and generated a viewport frame for each direction, as described in detail in Section \ref{sect:viewports}. For each viewport frame, a set of NSS features was computed exactly as by BRISQUE, then these features were averaged across all 18 viewports. We found that the shape and variance parameters of the GGD model from the two scales of the input frames are generally useful features for QP prediction across different databases, but did not use the shape parameter from the downscaled frames to match the features used in FOVQA. An SVR was then trained to map the selected 3 NSS features to the ground truth QP.

When predicting the QP maps for the quality fall-off model, for each viewport frame, we applied the previously learned SVR to the spatial shape and variance maps of SV-GGD, and the variance map of SV-GGD from the downscaled input, in a pixel-by-pixel manner in the feature map space, yielding a QP map for each viewport frame. A comparison of a ground truth QP map and a predicted QP map is shown in Fig. \ref{fig:QP_maps_comp}, where the predicted QP map is averaged across all the viewports of a foveated video. It may be seen that the predicted map follows the ground truth map, but noisier as could be expected.

Then, to compute the gradient information, we designed a set of filters for the FVQA problem by defining Gaussian smoothed derivatives in the radial direction in polar coordinates. There are two reasons for choosing Gaussian smoothed derivatives: i) The Gaussian smoothing reduces the effects of noise in the predicted QP maps, and ii) the variance of the Gaussian function allows us to capture gradient information at various scales of QP maps. The Gaussian smoothed derivatives in the radial direction are defined as:
\begin{equation}
\begin{split}
    G_{r}(\mathbf{r};\sigma_{G}) &= \frac{\partial}{\partial r}G(\mathbf{r};\sigma_{G})\\
    &= (\frac{\partial x}{\partial r}\frac{\partial}{\partial x}+\frac{\partial y}{\partial r}\frac{\partial}{\partial y})G(\mathbf{r};\sigma_{G})\\
    &= G_x(\mathbf{r};\sigma_{G})\cos\theta + G_y(\mathbf{r};\sigma_{G})\sin\theta,
\end{split}
\end{equation}
where $\mathbf{r}=(x,y)=(r\cos\theta,r\sin\theta)$, $r=||\mathbf{r}||$ is the radius with respect to the foveation point, $\theta$ is the polar angle, and $G(\mathbf{r};\sigma_{G})$ is a Gaussian function with variance $\sigma_{G}$:
\begin{equation}
    G(\mathbf{r};\sigma_{G}) = \frac{1}{\sigma_{G} \sqrt{2\pi}}\exp\left (-\frac{||\mathbf{r}||^{2}}{2\sigma_{G}^{2}}\right),
\end{equation}
$G_x$ and $G_y$ are orthogonal Gaussian derivatives in Cartesian coordinates:
\begin{equation}
    \begin{split}
        G_x(x,y) &= -\frac{x}{\sqrt{2\pi}\cdot \sigma^3}\cdot \exp^{-\frac{x^2+y^2}{2\sigma^2}}\\
        G_y(x,y) &= -\frac{y}{\sqrt{2\pi}\cdot \sigma^3}\cdot \exp^{-\frac{x^2+y^2}{2\sigma^2}}.
    \end{split}
\end{equation}
Then, the Gaussian derivative filters are applied to the predicted QP maps via convolution to obtain the gradient maps:
\begin{equation}
    D_{r}(\mathbf{r};\sigma_{G}) = QP(\mathbf{r})*G_{r}(\mathbf{r};\sigma_{G})
    \label{eq:radial_d}
\end{equation}
where $QP(\mathbf{r})$ is the QP map.

However, given the gradient of the QP map, it is difficult to quantify the overall amount of perceptual quality fall-off: the relation between the QPs and perceptual quality is complicated, so the quality fall-off distribution is a complex combination of fall-offs at different locations. Hence, we used a set of $\mathcal{K}_{fo}$ weighting patterns $w_{k}$ as in (\ref{eq:weightpatt}) to capture gradient information at different eccentricities, using these features to train an SVR. The feature extraction process is formulated as:
\begin{equation}
    f_{k,l}^{fo} = \sum_{\mathbf{x}} w_{k}(\mathbf{r})\cdot D_{r}(\mathbf{r};\sigma_{G}^{l}),
    \label{eq:fo_feats}
\end{equation}
where $\sigma_{G}^{l},\; l \in \{1,2,...,L\}$ is a set of variances controlling the scale of the gradient. In total, we extracted $\mathcal{K}_{fo}L$ fall-off features.

One alternative method is to first filter the QP maps using orthogonal Gaussian derivatives, then compute the root-squared-sum of the responses (gradient magnitude), as in many edge-detection related tasks \cite{dmarr1980,bovik_image}:
\begin{equation}
\begin{split}
    &D_{x}(\mathbf{r};\sigma_{G}) = QP(\mathbf{r})*G_{x}(\mathbf{r};\sigma_{G})\\
    &D_{y}(\mathbf{r};\sigma_{G}) = QP(\mathbf{r})*G_{y}(\mathbf{r};\sigma_{G})\\
    &D_{mag}(\mathbf{r};\sigma_{G}) = \sqrt{D_{x}^{2}(\mathbf{r};\sigma_{G})+D_{y}^{2}(\mathbf{r};\sigma_{G})}.
    \label{eq:mag_d}
\end{split}
\end{equation}

However, we use the filters ($G_{r}(\mathbf{r};\sigma_{G})$) due to the following reasons: i) variations of foveation distortions generally occur along the radial direction, hence it is desirable to take Gaussian smoothed derivatives along the radial direction to capture these changes. ii) When applied to radially symmetric maps, such as the ground truth QP maps (where $QP(\mathbf{r})$ can be written as $QP(r)$), the gradient map $D_{r}(\mathbf{r};\sigma_{G})$ in (\ref{eq:radial_d}) is the same as $D_{mag}(\mathbf{r};\sigma_{G})$ in (\ref{eq:mag_d}). iii) Most importantly, $D_{r}(\mathbf{r};\sigma_{G})$ is linearly related to $QP(\mathbf{r})$, so that noise in the predicted QP maps can be reduced via simple online updating, such as feature averaging. When applied to VQA problems, the spatial quality-aware features are generally applied on a frame-by-frame basis, then averaged across all frames. In FVQA, the per-frame space-variant statistics or predicted QP maps are generally noisy due to there being a limited number of samples used in local estimation, which will affect the accuracies of the extracted quality-aware features. However, the noise can be reduced by averaging across multiple frames since (\ref{eq:radial_d}) and (\ref{eq:fo_feats}) define a linear mapping between fall-off features and QP maps, and the noise-reducing property is preserved through the linearity.
\begin{comment}
\begin{table*}[t]
    \centering
    \caption{Choices of hyperparameters for the radial feature extraction. A value of 0 indicates the Gaussian function is a constant of value 0. A value of $\infty$ indicates the Gaussian function is a constant of value 1.}
    \begin{tabular}{ccccccccccc}
    \hline\hline
        k & 1 & 2 & 3 & 4 & 5 & 6 & 7 & 8 & 9 & 10 &11 &12 &13 &14 &15 \\\hline
        $r_{k}$ & $0$ & $9/6$ & $13/6$ & $17/6$ & $21/6$ & $25/6$ & $29/6$ & $33/6$ & $37/6$ & $\infty$ \\\hline
        $\sigma_{k,2}$ & $0$ & $5/6$ & $9/6$ & $13/6$ & $17/6$ & $21/6$ & $25/6$ & $29/6$ & $33/6$ & $37/6$ \\\hline\hline
    \end{tabular}
    \label{tab:sigma}
\end{table*}
\end{comment}
\begin{table*}[t]
    \centering
    \caption{Summary of selected parameter maps.}
    \resizebox{\textwidth}{!}{\begin{tabular}{c|c|c}
    \hline\hline
        Feature ID & Feature generation & Computation Procedure \\\hline
        $f_{1}$--$f_{2\mathcal{K}}$ & apply $w_{k}$ to the shape and variance maps & Fit SV-GGD to MSCN coefficients \\
        $f_{2\mathcal{K}+1}$--$f_{18\mathcal{K}}$ & apply $w_{k}$ to the shape, mean, left variance, and right variance maps & Fit SV-AGGD to pairwise products\\
        $f_{18\mathcal{K}+1}$--$f_{19\mathcal{K}}$ & apply $w_{k}$ to the variance map & Downscale the input, then fit SV-GGD to MSCN coefficients\\
        $f_{19\mathcal{K}+1}$--$f_{27\mathcal{K}}$ & apply $w_{k}$ to the left variance and right variance maps & Downscale then fit SV-AGGD to pairwise products\\
        $f_{27\mathcal{K}+1}$--$f_{27\mathcal{K}+\mathcal{K}_{fo}L}$ & apply $w_{k}$ to the gradients of the predicted QP map & Predict the QP map using NSS features and compute the gradient map\\\hline
        $f_{1}^{nvs}$--$f_{2\mathcal{K}}^{nvs}$ & apply $w_{k}$ to the shape and variance maps & Fit SV-GGD to MSCN coefficients from DFD$^{\dagger}$\\
        $f_{2\mathcal{K}+1}^{nvs}$--$f_{18\mathcal{K}}^{nvs}$ & apply $w_{k}$ to the shape, mean, left variance, and right variance maps & Fit SV-AGGD to pairwise products from DFD\\
        $f_{18\mathcal{K}+1}^{nvs}$--$f_{19\mathcal{K}^{nvs}}$ & apply $w_{k}$ to the variance map & Downscale the input, then fit SV-GGD to MSCN coefficients from DFD\\
        $f_{19\mathcal{K}+1}^{nvs}$--$f_{27\mathcal{K}}^{nvs}$ & apply $w_{k}$ to the left variance and right variance maps & Downscale then fit SV-AGGD to pairwise products from DFD\\\hline\hline
    \end{tabular}}\\
    \begin{flushleft}
    $^{\dagger}$ DFD: displaced frame differences.
    \end{flushleft}
    \label{tab:mtx_select}
\end{table*}

\begin{table*}[t]
    \centering
    \caption{Directions used in the evaluation framework.}
    \vspace{-2mm}
    \tiny
    \begin{tabular}{c|cccccc|cccccc|cccccc}
        \hline\hline
         Longitude & 0 & $\pi/6$ & $\pi/3$ & $\pi/2$ & $2\pi/3$ & $5\pi/6$ & 0 & $\pi/6$ & $\pi/3$ & $\pi/2$ & $2\pi/3$ & $5\pi/6$ & 0 & $\pi/6$ & $\pi/3$ & $\pi/2$ & $2\pi/3$ & $5\pi/6$  \\\hline
         Latitude & $\pi/4$ & $\pi/4$ & $\pi/4$ & $\pi/4$ & $\pi/4$ & $\pi/4$ & $\pi/2$ & $\pi/2$ & $\pi/2$ & $\pi/2$ & $\pi/2$ & $\pi/2$ & $3\pi/4$ & $3\pi/4$ & $3\pi/4$ & $3\pi/4$ & $3\pi/4$ & $3\pi/4$\\\hline\hline
    \end{tabular}
    \label{tab:dirs}
\end{table*}

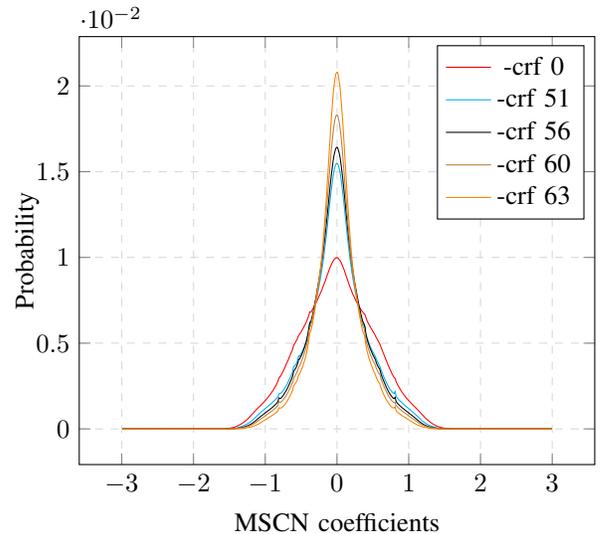
\begin{figure}[t]
    \centering
    \begin{tikzpicture}
        \begin{axis}[
            xlabel     = {MSCN coefficients},
            ylabel     = {Probability},
            grid       = major,
            grid style = {dashed,gray!30},
            y label style={at={(0.12,0.5)},anchor=south},
            cycle list={red,cyan,black,brown,orange}
        ]
            \addplot+[no markers] table[x=x,y=y] {figures/QP0.txt};
            \addplot+[no markers] table[x=x,y=y] {figures/QP1.txt};
            \addplot+[no markers] table[x=x,y=y] {figures/QP2.txt};
            \addplot+[no markers] table[x=x,y=y] {figures/QP3.txt};
            \addplot+[no markers] table[x=x,y=y] {figures/QP4.txt};
            \legend{-crf 0, -crf 51, -crf 56, -crf 60, -crf 63}
        \end{axis}
    \end{tikzpicture}
    \caption{Distribution of MSCN coefficients after displaced differencing frames. }
    \label{fig:disp_frame_diff}
\end{figure}

\section{The Space-Variant NVS Model} \label{sect:s4}
As first shown in \cite{opt_flow_st}, there are no general regularities of the statistics of motion vectors \cite{strred}. However, as shown in \cite{strred, speedqa}, high-quality video frame difference signals, following bandpass filtering and normalization by local contrast, obey reliable space-time statistical regularities. These regularities are impaired by distortions, and capturing measurements of these altered statistics yields important quality-aware information. An example of the distribution of MSCN coefficients after spatially displaced frame differencing is shown in Fig. \ref{fig:disp_frame_diff}

The spatial-temporal statistics of videos, also generally referred as Natural Video Statistics (NVS), are widely used in NR VQA models \cite{viideo,vbliinds,dashi2020}. In our proposed model, instead of using frame differences, we deploy the statistics of displaced differences of foveated frames \cite{lee2020} using SV-GGDs (\ref{eq:svggd}) and SV-AGGDs (\ref{eq:svaggd}).

We also applied the previously described neural noise model on the displaced frame differences for the same reasons:
\begin{equation}
    I_{k,d_{1},d_{2}}^{nvs}(i,j) = I_{k+1}(i,j) - I_{k}(i-d_{1},j-d_{2}) + \mathcal{W}_{nvs},
    \label{eq:nvs}
\end{equation}
where $(d_{1}, d_{2}) \in \{(0,1),(0,-1),(1,0),(-1,0)\}$, $I_{k}$ indicates the $k^{th}$ frame, and $\mathcal{W}_{nvs} \sim \mathcal{N}(0,w_{nvs})$. We then compute the MSCN coefficients as:
\begin{equation}
    \hat{I}_{k,d_{1},d_{2}}^{nvs} (i,j) = \frac{I_{k,d_{1},d_{2}}^{nvs}(i,j)-\mu_{k,d_{1},d_{2}}^{nvs} (i,j)}{\sigma_{k,d_{1},d_{2}}^{nvs} (i,j) + C_{nvs}}.
    \label{eq:MSCN_noise_nvs}
\end{equation}

The MSCN coefficients ($I_{k}^{nvs}$) and paired products of the MSCN coefficients (\ref{eq:paird_prod}) are modeled as obeying SV-GGDs and SV-AGGDs, respectively, where we have used col-located blocks of MSCN coefficients (or paired products) across all displacements ($d_{1},d_{2}$) to yield a single set of parameter maps to stabilize the statistics. Thus, we extracted $27\mathcal{K}$ features from the NVS model by the weighting patterns $w_{k}(\mathbf{r})$ to the extracted NVS feature maps, and trained another SVR to map these features to quality scores. Then the scores obtained from the spatial and temporal SVR outputs are combined using exponential weights:
\begin{equation}
    STScore = SScore^{\gamma}TScore^{1-\gamma},
\end{equation}
where $SScore$ and $TScore$ are the predictions from the two trained SVRs, and $\gamma$ weights their relative importances. We found that the combined scores accurately predict the ground truth quality scores when $\gamma$ lies within a range of [0.65,0.95]. We summarize all of the extracted features in TABLE \ref{tab:mtx_select}.
\section{Performance and Analysis}\label{sect:s5}
We use the newly built 2D LIVE-FBT-FCVR database to compare our proposed FOVQA model against a large variety of existing FIQA / FVQA algorithms. We first describe the evaluation framework under which all of the algorithms were compared. Four criteria were used: Pearson's linear correlation coefficient (PLCC), root mean squared error (RMSE), Spearman's rank order correlation coefficient (SROCC), and Kendall's rank order correlation coefficient (KROCC). Following usual practice \cite{sbrisque}, logistic non-linearity was applied to the predicted scores before computing PLCC and RMSE:
\begin{equation}
    Q(x) = \beta_{2} + \frac{\beta_{1}-\beta_{2}}{1+\exp(-\frac{x-\beta_{3}}{|\beta_4|})}.
    \label{eq:logistic_nonlin}
\end{equation}
Finally, we analyze the performance of the proposed FOVQA algorithm.

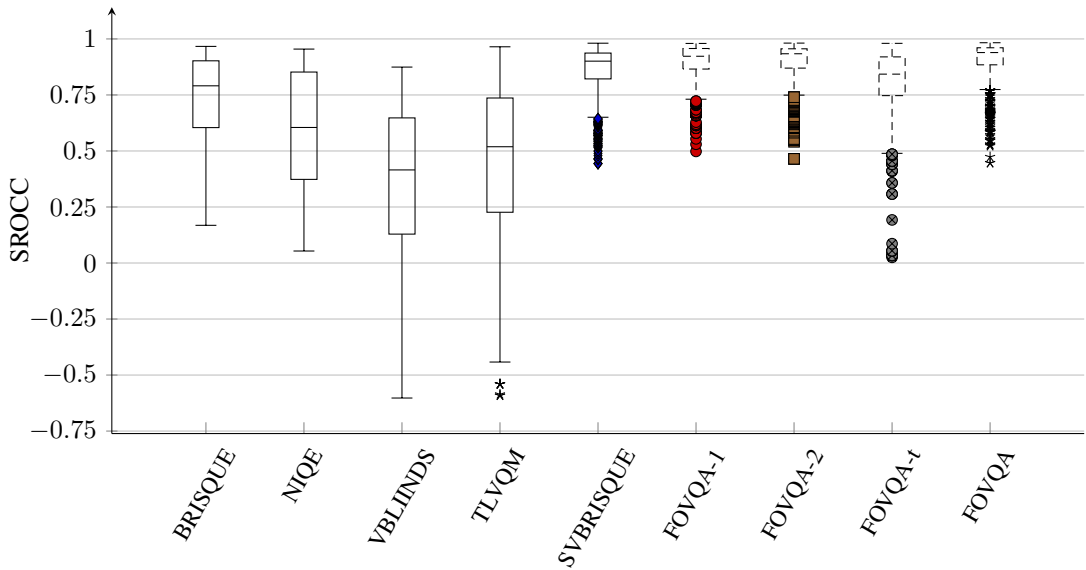
\begin{figure*}
    \centering
    \pgfplotstableread[col sep=comma]{figures/srocc.txt}\csvdata
    \pgfplotstabletranspose\datatransposed{\csvdata} 
    \begin{tikzpicture}
        \pgfplotsset{%
        width=.8\textwidth,
        height=.4\textwidth
        }
        \begin{axis}[
		boxplot/draw direction = y,
		axis x line* = bottom,
		axis y line = left,
		enlarge y limits,
		ymajorgrids,
		xtick = {3,6,9,12,15,18,21, 24, 27},
		xticklabel style = {align=center, font=\small, rotate=60},
		xticklabels = {BRISQUE, NIQE, VBLIINDS, TLVQM, SVBRISQUE, FOVQA-1, FOVQA-2, FOVQA-t, FOVQA},
		%xtick style = {draw=none}, % Hide tick line
		ylabel = {SROCC},
		ytick = {-1.0,-0.75,-0.5,-0.25,0,0.25,0.5,0.75,1.0}
	]
		\foreach \n in {1,...,9} {
			\addplot+[boxplot={draw position=\n*3},draw=black] table[y index=\n] {\datatransposed};
		}
		\end{axis}
    \end{tikzpicture}
    \caption{SROCC performance of NR IQA / VQA algorithms. FOVQA-1 shows the results when only radial basis features are used, FOVQA-2 shows the results when quality fall-off features are included, FOVQA-t shows the results when only temporal features are used, and finally FOVQA shows the performance when combining FOVQA-2 and temporal FOVQA-t using $\gamma=0.85$.}
    \label{fig:step_analysis}
\end{figure*}

\subsection{Evaluation Framework} \label{sect:viewports}
To recover the foveated experience and to better enable the more realistic comparison of algorithms, we simulate real-time foveation in the 2D LIVE-FBT-FCVR database \cite{yizetip2020} in the same way as \cite{sbrisque}, i.e. a set of 3D viewing directions were selected on the unit sphere, and uniformly distributed in terms of longitude and latitude, as shown in TABLE \ref{tab:dirs}. Along each viewing direction, a viewport video for each foveation distortion was generated at a resolution of 1024x1024 and a $90^{\circ}$ FOV. The compared VQA models are evaluated on these viewport videos. 

\begin{table}[t]
    \centering
    \scriptsize
    \caption{Comparison of VQA models on the 2D LIVE-FBT-FCVR database. The highest performances are boldfaced.}
    \resizebox{\columnwidth}{!}{\begin{tabular}{c|cccc}
    \hline\hline
         Methods & SROCC$\uparrow$ & KROCC$\uparrow$ & PLCC$\uparrow$ & RMSE$\downarrow$ \\\hline
         PSNR & 0.6954 & 0.5044 & 0.6941 & 7.09\\
         SSIM & 0.7191 & 0.5250 & 0.7260 & 6.77\\
         MSSSIM & 0.7243 & 0.5273 & 0.7288 & 6.75\\
         VIF & 0.8068 & 0.6182 & 0.8102 & 5.77\\
         SRRED & 0.7885 & 0.5873 & 0.7896 & 6.04\\
         STRRED & 0.7010 & 0.5182 & 0.6922 & 7.10\\
         SPEEDIQA & 0.7866 & 0.5872 & 0.7760 & 6.21\\
         SPEEDVQA & 0.6238 & 0.4539 & 0.6584 & 7.41\\
         FSIM & 0.7808 & 0.5850 & 0.7752 & 6.22\\
         FWQI & 0.7848 & 0.5909 & 0.7906 & 6.03\\
         FASSIM & 0.7418 & 0.5531 & 0.7573 & 6.43\\
         VMAF & 0.8103 & 0.6176 & 0.8047 & 5.84\\
         BRISQUE & 0.797$\pm$0.22 & 0.639$\pm$0.18 & 0.708$\pm$0.18 & 9.60$\pm$3.29\\
         NIQE & 0.605$\pm$0.32 & 0.457$\pm$0.24 & 0.675$\pm$0.31 & 6.47$\pm$2.27\\
         V-BLIINDS & 0.440$\pm$0.25 & 0.327$\pm$0.20 & 0.431$\pm$0.25 & 11.11$\pm$2.07\\
         TLVQM & 0.509$\pm$0.36 & 0.381$\pm$0.26 & 0.470$\pm$0.36 & 10.38$\pm$3.09\\
         SVBRISQUE & 0.900$\pm$0.11 & 0.736$\pm$0.12 & 0.884$\pm$0.10 & 6.91$\pm$2.53 \\
         FOVQA & \textbf{0.939$\pm$0.10} & \textbf{0.804$\pm$0.12} & \textbf{0.921$\pm$0.10} & \textbf{5.09$\pm$2.33}\\\hline\hline
    \end{tabular}}
    \label{tab:comparison}
\end{table}

\subsection{Choice of Hyperparameters in FOVQA}
We set the block size $P$ in both the SV-GGD and SV-AGGD models to 32. For the radial basis feature extraction in Section \ref{sect:weight_patt}, we set $w_{s}=0.01$ and $C_{s}=0.1$, and chose $\mathcal{K}=10$, achieving a fine granularity of features as compared to SVBRISQUE, where the values of $r_{k}$ are linearly spaced between 0 and 20, thus obtaining 270 features. The spread parameter $\sigma_{G_{RBF}}$ was simply set to the spacing between two adjacent $r_{k}$ values ($r_k-r_{k-1}$). 

For the perceptual quality fall-off model, we used $L=7$ by setting $\sigma_{G}^{l}\; \in \{0.5,0.75, 1.0,1.25, 1.5,1.75,2.0\}$ and set the values of $r_{k}$ linearly spaced between 0 and 13, since we have found that, outside this range, there is no QP fall-off on the videos in the 2D LIVE-FBT-FCVR database. We set $\mathcal{K}_{fo} = 6$, yielding 42 features. An SVR was trained on these overall 312 spatial features. 

For the NVS model, we similarly obtained 270 features, and another SVR was trained on these temporal features. When combining the spatial and temporal models, we set $\gamma=0.85$, and we show that varying $\gamma \in [0.65, 0.95]$ yields robust performance in Section \ref{sect:gam}.

\subsection{Comparison of Algorithms}
Among FR IQA algorithms, we included PSNR, SSIM \cite{ssim}, MS-SSIM \cite{msssim}, VIF \cite{vif}, S-RRED \cite{rred}, Speed-IQA \cite{speedqa}, FSIM \cite{fsim}, FWQI \cite{fwqi}, and FA-SSIM \cite{fassim}. These IQA algorithms were applied on every frame of the 18 viewports of every foveation compression distorted video, then were average pooled into a single final score. Among FR / RR VQA algorithms, we included ST-RRED \cite{strred}, Speed-VQA \cite{speedqa}, and VMAF \cite{vmaf}. These VQA algorithms were applied on each of the 18 viewport videos, and averaged into a single final score.

Among the compared NR VQA algorithms, we included BRISQUE \cite{brisque}, SVBRISQUE\cite{sbrisque}, NIQE \cite{niqe}, V-BLIINDS\cite{vbliinds}, TLVQM \cite{tlvqm}, and the proposed FOVQA algorithm. To implement BRISQUE, SVBRISQUE, V-BLIINDS, and TLVQM, we extracted and averaged features across every frame of each viewport video, obtaining one set of features for each foveated / distorted video, and trained an SVR to map features into quality scores. For NIQE, we followed the same procedure as the FR VQA models, by averaging quality scores obtained from each viewport frame and over all 18 viewport videos. 

For these NR algorithms, we ran the following procedure 1000 times and reported the median performance: first split the database into train ($80\%$) and test datasets ($20\%$) without content overlaps, then use a 4-fold cross validation to find the best hyperparameters for the SVR, finally train the SVR using the training set and record the performance on the test set.

The comparison of algorithms on the 2D LIVE-FBT-FCVR database is shown in TABLE \ref{tab:comparison}, where the predictions from FR algorithms are compared against Difference Mean Opinion Scores (DMOS), while the NR algorithms were trained on and compared against Mean Opinion Scores (MOS), following common practice. It may be readily observed that FOVQA algorithm was able to achieve a significant leap in SOTA performance, reaching a rank-order correlation of 0.939 against the human judgments. 
\begin{comment}
In addition, FOVQA provides significantly better stability of predictions, yielding much smaller variances in performance across the 1000 iterations of the train-test procedure.
\end{comment}

\begin{table}[t]
    \centering
    \scriptsize
    \caption{Ablation study of the proposed FOVQA algorithm. FOVQA-1 shows the results when only the radial basis features are used, FOVQA-2 shows the results when quality fall-off features are also applied, FOVQA-t shows the results when only temporal features are used, and FOVQA shows the combined performance when $\gamma=0.85$. Best performances are boldfaced.}
    \resizebox{\columnwidth}{!}{\begin{tabular}{c|cccc}
    \hline\hline
         Methods & SROCC$\uparrow$ & KROCC$\uparrow$ & PLCC$\uparrow$ & RMSE$\downarrow$ \\\hline
         FOVQA-1 & {0.922$\pm$0.10} & {0.778$\pm$0.11} & {0.904$\pm$0.10} & {5.78$\pm$1.89}\\\hline
         FOVQA-2 & {0.934$\pm$0.10} & {0.795$\pm$0.12} & {0.916$\pm$0.10} & {5.29$\pm$2.21}\\\hline
         FOVQA-t & {0.843$\pm$0.16} & {0.673$\pm$0.15} & {0.846$\pm$0.14} & {7.20$\pm$2.20}\\\hline
         FOVQA & \textbf{0.939$\pm$0.10} & \textbf{0.804$\pm$0.12} & \textbf{0.921$\pm$0.10} & \textbf{5.09$\pm$2.33}\\\hline\hline
    \end{tabular}}
    \label{tab:ablation}
\end{table}

\subsection{Ablation Study}
We conducted an ablation study by performing a step-by-step analysis of the performance of FOVQA, as shown in Fig. \ref{fig:step_analysis} and TABLE \ref{tab:ablation}. It may be observed that both the spatial and temporal FOVQA models outperformed other prior existing models by a large margin. In addition, it may be seen that the perceptual quality fall-off features provide complementary information to the other NSS features.

\begin{table}[t]
    \centering
    \scriptsize
    \caption{Robustness of performance against different values of $\gamma$. Best performances are boldfaced.}
    \resizebox{\columnwidth}{!}{\begin{tabular}{c|cccc}
    \hline\hline
         $\gamma$ & SROCC$\uparrow$ & KROCC$\uparrow$ & PLCC$\uparrow$ & RMSE$\downarrow$ \\\hline
         0.65 & {0.934$\pm$0.11} & {0.792$\pm$0.12} & \textbf{0.924$\pm$0.11} & {5.12$\pm$2.42}\\\hline
         0.7 & {0.936$\pm$0.10} & {0.795$\pm$0.12} & {0.922$\pm$0.05} & {5.08$\pm$2.41}\\\hline
         0.75 & {0.938$\pm$0.10} & {0.801$\pm$0.12} & {0.922$\pm$0.10} & {5.07$\pm$2.39}\\\hline
         0.8 & {0.938$\pm$0.10} & {0.801$\pm$0.12} & {0.921$\pm$0.10} & \textbf{5.06$\pm$2.37}\\\hline
         0.85 & \textbf{0.939$\pm$0.10} & \textbf{0.804$\pm$0.12} & {0.921$\pm$0.10} & {5.09$\pm$2.33}\\\hline
         0.9 & \textbf{0.939$\pm$0.10} & {0.801$\pm$0.12} & {0.920$\pm$0.10} & {5.17$\pm$2.29}\\\hline
         0.95 & {0.937$\pm$0.10} & {0.801$\pm$0.12} & {0.919$\pm$0.10} & {5.22$\pm$2.25}\\\hline\hline
    \end{tabular}}
    \label{tab:gamma}
\end{table}

\subsection{Rubustness of \texorpdfstring{$\gamma$}{Lg}}
\label{sect:gam}
We studied the performance obtained by combining the spatial and temporal outputs of the corresponding SVRs using different values of $\gamma$, as shown in TABLE \ref{tab:gamma}. For better comparison, we used a fixed set of 1000 train-test splits. It may be seen that the performance of FOVQA is robust against the choice of $\gamma$, but skewed towards spatial predictions. 

\section{Conclusion and Future Work}\label{sect:s6}
We extended traditional natural scene statistics (NSS) and natural video statistics (NVS) models into space-variant NSS / NVS models suitable for foveated compression scenarios, and proposed a new foveated video quality assessment algorithm, called FOVQA, that is based on space-variant generalized Gaussian distributions (SV-GGD) and space-variant asynchronous generalized Gaussian distributions (SV-AGGD). The new FOVQA algorithm has several distinguishing elements, including the use of parametric space-variant statistical models, new feature extraction / pooling schemes, and unique features that model the rapidity of quality fall-off of foveation distortions, which can significantly degrade perceptual quality. We conducted ablation studies to show that the various 'quality-aware' features we utilize in FOVQA indeed provide beneficial, but not redundant information towards predicting perceptual quality.

One possible future direction is to reduce the time complexity of parameter estimation of the space-variant models. While the computational complexity of FOVQA is already significantly reduced by using non-overlapping local patches, computing the parameter maps still requires significant computation. A light-weight FVQA algorithm would be desirable and beneficial for real-time foveated VR video quality assessment applications.

\section*{Acknowledgment}

The authors thank Facebook Technologies for the fruitful discussions and for supporting this research. The authors would also like to thank Li-Heng Chen for sharing computational resources.

% Can use something like this to put references on a page
% by themselves when using endfloat and the captionsoff option.
\ifCLASSOPTIONcaptionsoff
  \newpage
\fi

% trigger a \newpage just before the given reference
% number - used to balance the columns on the last page
% adjust value as needed - may need to be readjusted if
% the document is modified later
%\IEEEtriggeratref{8}
% The "triggered" command can be changed if desired:
%\IEEEtriggercmd{\enlargethispage{-5in}}

% references section

% can use a bibliography generated by BibTeX as a .bbl file
% BibTeX documentation can be easily obtained at:
% http://mirror.ctan.org/biblio/bibtex/contrib/doc/
% The IEEEtran BibTeX style support page is at:
% http://www.michaelshell.org/tex/ieeetran/bibtex/
\bibliography{Transactions-Bibliography/IEEEexample.bib}{}

% Generated by IEEEtran.bst, version: 1.14 (2015/08/26)
\begin{thebibliography}{10}
\providecommand{\url}[1]{#1}
\csname url@samestyle\endcsname
\providecommand{\newblock}{\relax}
\providecommand{\bibinfo}[2]{#2}
\providecommand{\BIBentrySTDinterwordspacing}{\spaceskip=0pt\relax}
\providecommand{\BIBentryALTinterwordstretchfactor}{4}
\providecommand{\BIBentryALTinterwordspacing}{\spaceskip=\fontdimen2\font plus
\BIBentryALTinterwordstretchfactor\fontdimen3\font minus
  \fontdimen4\font\relax}
\providecommand{\BIBforeignlanguage}[2]{{%
\expandafter\ifx\csname l@#1\endcsname\relax
\typeout{** WARNING: IEEEtran.bst: No hyphenation pattern has been}%
\typeout{** loaded for the language `#1'. Using the pattern for}%
\typeout{** the default language instead.}%
\else
\language=\csname l@#1\endcsname
\fi
#2}}
\providecommand{\BIBdecl}{\relax}
\BIBdecl

\bibitem{Ryoo2016}
J.~Ryoo, K.~Yun, D.~Samaras, S.~R. Das, and G.~Zelinsky, ``Design and
  evaluation of a foveated video streaming service for commodity client
  devices,'' in \emph{Proceedings of the 7th International Conference on
  Multimedia Systems}, ser. MMSys '16.\hskip 1em plus 0.5em minus 0.4em\relax
  New York, NY, USA: Association for Computing Machinery, 2016.

\bibitem{Romero2018}
\BIBentryALTinterwordspacing
M.~F. Romero-Rond\'{o}n, L.~Sassatelli, F.~Precioso, and R.~Aparicio-Pardo,
  ``Foveated streaming of virtual reality videos,'' in \emph{Proceedings of the
  9th ACM Multimedia Systems Conference}, ser. MMSys '18.\hskip 1em plus 0.5em
  minus 0.4em\relax New York, NY, USA: Association for Computing Machinery,
  2018, p. 494–497. [Online]. Available:
  \url{https://doi.org/10.1145/3204949.3208114}
\BIBentrySTDinterwordspacing

\bibitem{Kim2018}
\BIBentryALTinterwordspacing
H.~Kim, J.~Yang, M.~Choi, J.~Lee, S.~Yoon, Y.~Kim, and W.~Park, ``Eye tracking
  based foveated rendering for 360 vr tiled video,'' in \emph{Proceedings of
  the 9th ACM Multimedia Systems Conference}, ser. MMSys '18.\hskip 1em plus
  0.5em minus 0.4em\relax New York, NY, USA: Association for Computing
  Machinery, 2018, p. 484–486. [Online]. Available:
  \url{https://doi.org/10.1145/3204949.3208111}
\BIBentrySTDinterwordspacing

\bibitem{Illahi2020}
\BIBentryALTinterwordspacing
G.~K. Illahi, T.~V. Gemert, M.~Siekkinen, E.~Masala, A.~Oulasvirta, and
  A.~Yl\"{a}-J\"{a}\"{a}ski, ``Cloud gaming with foveated video encoding,''
  \emph{ACM Trans. Multimedia Comput. Commun. Appl.}, vol.~16, no.~1, Feb.
  2020. [Online]. Available: \url{https://doi.org/10.1145/3369110}
\BIBentrySTDinterwordspacing

\bibitem{yizetip2020}
Y.~Jin, M.~Chen, T.~Goodall, A.~Patney, and A.~C. Bovik, ``Subjective and
  objective quality assessment of 2d and 3d foveated video compression in
  virtual reality,'' \emph{IEEE Transactions on Image Processing}, submitted
  for publication.

\bibitem{dashi2020}
X.~Yu, N.~Birkbeck, Y.~Wang, C.~G. Bampis, B.~Adsumilli, and A.~C. Bovik,
  ``Predicting the quality of compressed videos with pre-existing
  distortions,'' \emph{IEEE Transactions on Image Processing}, submitted for
  publication.

\bibitem{lee2020}
D.~Lee, H.~Ko, J.~Kim, and A.~C. Bovik, ``On the space-time statistics of
  motion pictures,'' in \emph{Journal of Vision}, submitted for publication.

\bibitem{ssim}
{Z. Wang}, A.~C. {Bovik}, H.~R. {Sheikh}, and E.~P. {Simoncelli}, ``Image
  quality assessment: from error visibility to structural similarity,''
  \emph{IEEE Transactions on Image Processing}, vol.~13, no.~4, pp. 600--612,
  2004.

\bibitem{msssim}
Z.~{Wang}, E.~P. {Simoncelli}, and A.~C. {Bovik}, ``Multiscale structural
  similarity for image quality assessment,'' in \emph{The Thrity-Seventh
  Asilomar Conference on Signals, Systems Computers, 2003}, vol.~2, 2003, pp.
  1398--1402 Vol.2.

\bibitem{ifc}
H.~R. {Sheikh}, A.~C. {Bovik}, and G.~{de Veciana}, ``An information fidelity
  criterion for image quality assessment using natural scene statistics,''
  \emph{IEEE Transactions on Image Processing}, vol.~14, no.~12, pp.
  2117--2128, 2005.

\bibitem{vif}
H.~R. {Sheikh} and A.~C. {Bovik}, ``Image information and visual quality,''
  \emph{IEEE Transactions on Image Processing}, vol.~15, no.~2, pp. 430--444,
  2006.

\bibitem{gsm}
M.~J. Wainwright and E.~P. Simoncelli, ``Scale mixtures of gaussians and the
  statistics of natural images,'' ser. NIPS'99.\hskip 1em plus 0.5em minus
  0.4em\relax Cambridge, MA, USA: MIT Press, 1999, p. 855–861.

\bibitem{fsim}
L.~{Zhang}, L.~{Zhang}, X.~{Mou}, and D.~{Zhang}, ``Fsim: A feature similarity
  index for image quality assessment,'' \emph{IEEE Transactions on Image
  Processing}, vol.~20, no.~8, pp. 2378--2386, 2011.

\bibitem{vqm}
M.~H. {Pinson} and S.~{Wolf}, ``A new standardized method for objectively
  measuring video quality,'' \emph{IEEE Transactions on Broadcasting}, vol.~50,
  no.~3, pp. 312--322, 2004.

\bibitem{movie}
K.~{Seshadrinathan} and A.~C. {Bovik}, ``A model of neuronal responses in
  visual area mt,'' \emph{Vision Research}, vol.~19, no.~2, pp. 335--350, 1998.

\bibitem{areamt}
E.~P. Simoncelli and D.~J. Heeger, ``{A Model of Neural Responses in Area
  MT},'' \emph{Vision research}, vol.~38, no.~5, pp. 743--761, 1998.

\bibitem{vmaf}
\BIBentryALTinterwordspacing
Z.~Li, A.~Aaron, A.~Moorthy, and M.~Manohara. Toward a practical perceptual
  video quality metric. [Online]. Available:
  \url{http://techblog.netflix.com/2016/06/toward-practical-perceptual-video.html}
\BIBentrySTDinterwordspacing

\bibitem{dlm}
S.~{Li}, F.~{Zhang}, L.~{Ma}, and K.~N. {Ngan}, ``Image quality assessment by
  separately evaluating detail losses and additive impairments,'' \emph{IEEE
  Transactions on Multimedia}, vol.~13, no.~5, pp. 935--949, 2011.

\bibitem{rred}
R.~{Soundararajan} and A.~C. {Bovik}, ``Rred indices: Reduced reference
  entropic differencing for image quality assessment,'' \emph{IEEE Transactions
  on Image Processing}, vol.~21, no.~2, pp. 517--526, 2012.

\bibitem{strred}
------, ``Video quality assessment by reduced reference spatio-temporal
  entropic differencing,'' \emph{IEEE Transactions on Circuits and Systems for
  Video Technology}, vol.~23, no.~4, pp. 684--694, 2013.

\bibitem{speedqa}
C.~G. {Bampis}, P.~{Gupta}, R.~{Soundararajan}, and A.~C. {Bovik}, ``Speed-qa:
  Spatial efficient entropic differencing for image and video quality,''
  \emph{IEEE Signal Processing Letters}, vol.~24, no.~9, pp. 1333--1337, 2017.

\bibitem{brisque}
A.~{Mittal}, A.~K. {Moorthy}, and A.~C. {Bovik}, ``No-reference image quality
  assessment in the spatial domain,'' \emph{IEEE Transactions on Image
  Processing}, vol.~21, no.~12, pp. 4695--4708, 2012.

\bibitem{niqe}
A.~{Mittal}, R.~{Soundararajan}, and A.~C. {Bovik}, ``Making a “completely
  blind” image quality analyzer,'' \emph{IEEE Signal Processing Letters},
  vol.~20, no.~3, pp. 209--212, 2013.

\bibitem{svr}
\BIBentryALTinterwordspacing
C.~C. Chang and C.~J. Lin. Libsvm: A library for support vector machines.
  [Online]. Available: \url{http://www.csie.ntu.edu.tw/\~cjlin/libsvm/}
\BIBentrySTDinterwordspacing

\bibitem{ilniqe}
L.~{Zhang}, L.~{Zhang}, and A.~C. {Bovik}, ``A feature-enriched completely
  blind image quality evaluator,'' \emph{IEEE Transactions on Image
  Processing}, vol.~24, no.~8, pp. 2579--2591, 2015.

\bibitem{vbliinds}
M.~A. {Saad}, A.~C. {Bovik}, and C.~{Charrier}, ``Blind prediction of natural
  video quality,'' \emph{IEEE Transactions on Image Processing}, vol.~23,
  no.~3, pp. 1352--1365, 2014.

\bibitem{nvs}
\BIBentryALTinterwordspacing
D.~W. Dong and J.~J. Atick, ``Statistics of natural time-varying images,''
  \emph{Network: Computation in Neural Systems}, vol.~6, no.~3, pp. 345--358,
  1995. [Online]. Available: \url{https://doi.org/10.1088/0954-898X_6_3_003}
\BIBentrySTDinterwordspacing

\bibitem{viideo}
A.~{Mittal}, M.~A. {Saad}, and A.~C. {Bovik}, ``A completely blind video
  integrity oracle,'' \emph{IEEE Transactions on Image Processing}, vol.~25,
  no.~1, pp. 289--300, 2016.

\bibitem{tlvqm}
J.~{Korhonen}, ``Two-level approach for no-reference consumer video quality
  assessment,'' \emph{IEEE Transactions on Image Processing}, vol.~28, no.~12,
  pp. 5923--5938, 2019.

\bibitem{cvd2014}
M.~{Nuutinen}, T.~{Virtanen}, M.~{Vaahteranoksa}, T.~{Vuori}, P.~{Oittinen},
  and J.~{Häkkinen}, ``Cvd2014—a database for evaluating no-reference video
  quality assessment algorithms,'' \emph{IEEE Transactions on Image
  Processing}, vol.~25, no.~7, pp. 3073--3086, 2016.

\bibitem{konvid1k}
V.~{Hosu}, F.~{Hahn}, M.~{Jenadeleh}, H.~{Lin}, H.~{Men}, T.~{Szirányi},
  S.~{Li}, and D.~{Saupe}, ``The konstanz natural video database (konvid-1k),''
  in \emph{2017 Ninth International Conference on Quality of Multimedia
  Experience (QoMEX)}, 2017, pp. 1--6.

\bibitem{livequalcomm}
D.~{Ghadiyaram}, J.~{Pan}, A.~C. {Bovik}, A.~K. {Moorthy}, P.~{Panda}, and
  K.~{Yang}, ``In-capture mobile video distortions: A study of subjective
  behavior and objective algorithms,'' \emph{IEEE Transactions on Circuits and
  Systems for Video Technology}, vol.~28, no.~9, pp. 2061--2077, 2018.

\bibitem{fwqi}
\BIBentryALTinterwordspacing
Z.~Wang, A.~C. Bovik, L.~Lu, and J.~L. Kouloheris, ``{Foveated wavelet image
  quality index},'' in \emph{Applications of Digital Image Processing XXIV},
  A.~G. Tescher, Ed., vol. 4472, International Society for Optics and
  Photonics.\hskip 1em plus 0.5em minus 0.4em\relax SPIE, 2001, pp. 42 -- 52.
  [Online]. Available: \url{https://doi.org/10.1117/12.449797}
\BIBentrySTDinterwordspacing

\bibitem{csf}
\BIBentryALTinterwordspacing
W.~S. Geisler and J.~S. Perry, ``{Real-time foveated multiresolution system for
  low-bandwidth video communication},'' in \emph{Human Vision and Electronic
  Imaging III}, B.~E. Rogowitz and T.~N. Pappas, Eds., vol. 3299, International
  Society for Optics and Photonics.\hskip 1em plus 0.5em minus 0.4em\relax
  SPIE, 1998, pp. 294 -- 305. [Online]. Available:
  \url{https://doi.org/10.1117/12.320120}
\BIBentrySTDinterwordspacing

\bibitem{vdntm}
A.~B. {Watson}, G.~Y. {Yang}, J.~A. {Solomon}, and J.~{Villasenor},
  ``Visibility of wavelet quantization noise,'' \emph{IEEE Transactions on
  Image Processing}, vol.~6, no.~8, pp. 1164--1175, 1997.

\bibitem{fpsnr}
{Sanghoon Lee}, M.~S. {Pattichis}, and A.~C. {Bovik}, ``Foveated video quality
  assessment,'' \emph{IEEE Transactions on Multimedia}, vol.~4, no.~1, pp.
  129--132, 2002.

\bibitem{fassim}
S.~{Rimac-Drlje}, G.~{Martinović}, and B.~{Zovko-Cihlar}, ``Foveation-based
  content adaptive structural similarity index,'' in \emph{2011 18th
  International Conference on Systems, Signals and Image Processing}, 2011, pp.
  1--4.

\bibitem{fovcsf}
\BIBentryALTinterwordspacing
S.~Rimac-Drlje, M.~Vranje{\v{s}}, and D.~{\v{Z}}agar, ``Foveated mean squared
  error---a novel video quality metric,'' \emph{Multimedia Tools and
  Applications}, vol.~49, no.~3, pp. 425--445, Sep 2010. [Online]. Available:
  \url{https://doi.org/10.1007/s11042-009-0442-1}
\BIBentrySTDinterwordspacing

\bibitem{sbrisque}
Y.~Jin, T.~Goodall, A.~Patney, and A.~C. Bovik, ``A natural scene statistics
  based foveated video quality assessment model,'' \emph{IEEE International
  Conference on Image Processing}, submitted for publication.

\bibitem{ruderman1994}
{D. L. Ruderman}, ``The statistics of natural images,'' \emph{Network:
  Computation in Neural Systems}, vol.~5, no.~4, pp. 517--548, 1994.

\bibitem{sheikh2006}
H.~R. {Sheikh}, M.~F. {Sabir}, and A.~C. {Bovik}, ``A statistical evaluation of
  recent full reference image quality assessment algorithms,'' \emph{IEEE
  Transactions on Image Processing}, vol.~15, no.~11, pp. 3440--3451, 2006.

\bibitem{moorthy2010}
A.~K. {Moorthy} and A.~C. {Bovik}, ``Statistics of natural image distortions,''
  in \emph{2010 IEEE International Conference on Acoustics, Speech and Signal
  Processing}, 2010, pp. 962--965.

\bibitem{sharifi1995}
K.~{Sharifi} and A.~{Leon-Garcia}, ``Estimation of shape parameter for
  generalized gaussian distributions in subband decompositions of video,''
  \emph{IEEE Transactions on Circuits and Systems for Video Technology},
  vol.~5, no.~1, pp. 52--56, 1995.

\bibitem{livevqa}
K.~{Seshadrinathan}, R.~{Soundararajan}, A.~C. {Bovik}, and L.~K. {Cormack},
  ``Study of subjective and objective quality assessment of video,'' \emph{IEEE
  Transactions on Image Processing}, vol.~19, no.~6, pp. 1427--1441, 2010.

\bibitem{prafulxray}
P.~{Gupta}, J.~{Glover}, N.~{Paulter}, and A.~{Bovik}, ``Studying the
  statistics of natural x-ray pictures,'' \emph{Journal of Testing and
  Evaluation}, vol.~46, pp. 1478--1488, 2018.

\bibitem{dmarr1980}
D.~{Marr} and E.~{Hildbeth}, ``Theory of edge detection,'' \emph{Proc. R. Soc.
  Lond. B.}, vol. 207, pp. 187--217, 1980.

\bibitem{bovik_image}
A.~C. Bovik, \emph{The Essential Guide to Image Processing}.\hskip 1em plus
  0.5em minus 0.4em\relax 6277 Sea Harbor Drive Orlando, FL, United States:
  Academic Press, Inc., 2009.

\bibitem{opt_flow_st}
S.~{Roth} and M.~J. {Black}, ``On the spatial statistics of optical flow,'' in
  \emph{Tenth IEEE International Conference on Computer Vision (ICCV'05) Volume
  1}, vol.~1, 2005, pp. 42--49 Vol. 1.

\end{thebibliography}
\bibliographystyle{IEEEtran}
\end{document}